\documentclass[aps, prb, reprint, superscriptaddress, notitlepage, letterpaper,  10pt, floatfix, showpacs, longbibliography, balancelastpage]{revtex4-2}
\pdfoutput=1
\usepackage{amssymb, graphicx, makecell, color, xcolor, amsmath, bm, url, float, mathrsfs, braket, bbold, psfrag, dcolumn, enumitem, array, tikz, algpseudocode, setspace, physics, adjustbox, tabularx, ragged2e, booktabs, silence, flushend}
\usetikzlibrary{quantikz2}
\usepackage[normalem]{ulem}
\usepackage[mathlines]{lineno}
\usepackage[section]{placeins}
\usepackage[linesnumbered, ruled, vlined]{algorithm2e}
\usepackage[colorlinks = True, citecolor = blue, linkcolor = blue, anchorcolor=red, pdftoolbar = false, bookmarks = false]{hyperref}
\usepackage[utf8]{inputenc}
\usepackage{orcidlink}

\def\bea{\begin{eqnarray}}
\def\eea{\end{eqnarray}}
\def\bal{\begin{aligned}}
\def\eal{\end{aligned}}
\begin{document}
\title{Efficient Estimation and Sequential Optimization of Cost Functions in Variational Quantum Algorithms}
\author{Muhammad Umer\orcidlink{0000-0002-1941-1833}}
\email{umer@u.nus.edu}
\affiliation{Centre for Quantum Technologies, National University of Singapore, 3 Science Drive 2, Singapore 117543}
\author{Eleftherios Mastorakis\orcidlink{0009-0005-3546-2568}}
\affiliation{School of Electrical and Computer Engineering, Technical University of Crete, Chania, Greece 73100}
\author{Dimitris G. Angelakis\orcidlink{0000-0001-6763-6060}}
\email{dimitris.angelakis@gmail.com}
\affiliation{Centre for Quantum Technologies, National University of Singapore, 3 Science Drive 2, Singapore 117543}
\affiliation{School of Electrical and Computer Engineering, Technical University of Crete, Chania, Greece 73100}
\affiliation{AngelQ Quantum Computing, 531A Upper Cross Street, \#04-95 Hong Lim Complex, Singapore 051531}
\date{\today}

\begin{abstract}
Classical optimization is a cornerstone of the success of variational quantum algorithms, which often require determining the derivatives of the cost function relative to variational parameters. The computation of the cost function and its derivatives, coupled with their effective utilization, facilitates faster convergence by enabling smooth navigation through complex landscapes, ensuring the algorithm\textquotesingle{s} success in addressing challenging variational problems. In this work, we introduce a novel optimization methodology that conceptualizes the parameterized quantum circuit as a weighted sum of distinct unitary operators, enabling the cost function to be expressed as a sum of multiple terms. This representation facilitates the efficient evaluation of nonlocal characteristics of cost functions, as well as their arbitrary derivatives. The optimization protocol then utilizes the nonlocal information on the cost function to facilitate a more efficient navigation process, ultimately enhancing the performance in the pursuit of optimal solutions. We utilize this methodology for two distinct cost functions. The first is the squared residual of the variational state relative to a target state, which is subsequently employed to examine the nonlinear dynamics of fluid configurations governed by the one-dimensional Burgers\textquotesingle{} equation. The second cost function is the expectation value of an observable, which is later utilized to approximate the ground state of the nonlinear Schr\"{o}dinger equation. Our findings reveal substantial enhancements in convergence speed and accuracy relative to traditional optimization methods, even within complex, high-dimensional landscapes. Our work contributes to the advancement of optimization strategies for variational quantum algorithms, establishing a robust framework for addressing a range of computationally intensive problems across numerous applications.
\end{abstract}
\maketitle

\section{ Introduction }
\label{Sec:Introduction}

\par Quantum computation has attracted significant attention in recent decades due to its potential to efficiently address a range of classically intractable problems. In this regard, various quantum algorithms, such as Shor\textquotesingle{s} \cite{Shor1994}, Grover\textquotesingle{s} \cite{Grover1996}, Harrow-Hassidim-Lloyd\textquotesingle{s} (HHL) \cite{Harrow2009}, and quantum simulation \cite{Abrams1997, *Abrams1999} algorithms, have been developed to leverage quantum devices for solving complex problems across a range of domains with diverse practical applications. These algorithms, however, require fault-tolerant quantum computation \cite{Khodjasteh2005, Temme2017, Endo2018, Kim2023}, which remains beyond the capabilities of present-day Noisy Intermediate-Scale Quantum (NISQ) devices. One of the prominent areas of research for capitalizing NISQ devices is the development of variational quantum algorithms (VQAs) \cite{Cerezo2021, Bharti2022}, with applications spanning disciplines such as quantum chemistry \cite{Peruzzo2014, Wecker2015, OMalley2016, Kandala2017, Hempel2018, Ganzhorn2019, Quantum2020}, finance \cite{Cong2024, Huber2024, Sarma2024}, fluid dynamics \cite{Lubasch2020, Jaksch2023, Umer2024}, quantum dynamics \cite{Cirstoiu2020, Lin2021, Linteau2024}, optimization problems \cite{Farhi2014, Pagano2020, Tan2021, Zhu2022}, and machine learning \cite{Benedetti2019, Zhu2019, Tangpanitanon2020}.

\par Variational quantum algorithms (VQAs) employ a hybrid quantum-classical approach, utilizing a quantum device to construct approximate solutions to problems using parameterized quantum circuits (PQCs), whose parameters are optimized on a classical computer \cite{Peruzzo2014, Kandala2017, McClean2016}. A PQC generates a quantum state, enabling the evaluation of various quantities, such as the expectation value of an observable or the fidelity of the variational state relative to a target state. These quantities serve as the cost function for the classical optimization process, with their minimum or maximum characterizing the optimal parameters that represent the solution to the problem. This seemingly simple idea encounters numerous challenges that obstruct the functionality of VQAs and limit their scalability to real-world applications. One of the bottlenecks is the efficient classical optimization of variational parameters \cite{Mcclean2018, Bittel2021, Fontana2022, Anschuetz2022, Larocca2022, *Larocca2024, Qi2023, Qi2023}, a crucial step in the successful implementation of VQAs. Here, challenges such as local minima \cite{Bittel2021, Fontana2022, Anschuetz2022} and barren plateaus \cite{Mcclean2018, Larocca2022, *Larocca2024, Qi2023} significantly impact the efficacy of classical optimization protocols.

\par The classical optimization process in VQAs employs either gradient-based \cite{Broyden1970, *Fletcher1970, *Goldfarb1970, *Shanno1970, Kraft1988, Kingma2014} or gradient-free \cite{Powell1994, *Powell1998, *Powell2007, Spall1987, *Spall1992, Nelder1965} approaches to determine the optimal parameters of the PQC. The gradient-based methods rely on the computation of gradients and/or higher-order derivatives of the cost function relative to the parameters subject to optimization. These gradients are evaluated by employing parameter-shift rule \cite{Schuld2019, Crooks2019, Mari2021, Banchi2021, Hubregtsen2022} and analytical techniques \cite{Mitarai2019, Harrow2021}, providing the direction and magnitude of change for iterative parameter updates. On the contrary, gradient-free methods do not rely on any derivatives; instead, they explore the parameter space using heuristic or systematic strategies, such as random search and sequential testing \cite{Powell1994, *Powell1998, *Powell2007, Nelder1965}, to identify the optimal parameters. Most optimization protocols, including gradient-based methods such as gradient descent, ADAM \cite{Kingma2014}, BFGS \cite{Broyden1970, *Fletcher1970, *Goldfarb1970, *Shanno1970}, and SLSQP \cite{Kraft1988, Virtanen2020}, as well as gradient-free algorithms like COBYLA \cite{Powell1994, *Powell1998, *Powell2007}, SPSA \cite{Spall1987, *Spall1992}, and Nelder-Mead \cite{Nelder1965}, have problem-independent structures and often encounter challenges in converging to the global optimum of the cost function landscape. However, problem-dependent formulations, such as Rotosolve \cite{Ostaszewski2021} and sequential techniques \cite{Parrish2019, Nakanishi2020}, leverage system-specific information to guide the optimization process, facilitating more efficient parameter updates and enhancing the likelihood of convergence to the global optimum. Our work expands on these advancements and presents a VQA-tailored approach to efficiently estimate the cost function and its arbitrary derivatives and improve the parameter optimization process to enhance the performance of VQAs.

\par In this article, we propose a sequential optimization method specifically designed for variational quantum algorithms. Our approach involves expressing the parameterized quantum circuit (PQC) as a weighted sum of distinct unitary operators, enabling the cost function to be represented as a linear combination of multiple terms. This formulation simplifies the optimization task to determining the optimal weights for the unitary operators in the PQC.
This representation of the cost function enables analysis across the domain of an individual parameter, $\lambda_{j} \in [-\pi, \pi)$, facilitates explicit updates to variational parameters, and allows for the computation of arbitrary derivatives without additional quantum resources. Moreover, our approach is versatile and can be applied to a variety of cost functions commonly considered in VQA applications. In this work, we focus on two distinct cost functions: the squared residual of the variational state with respect to a target state, referred to as the squared residual cost function, and the expectation value of an observable, termed expectation cost function. We apply our approach to one-dimensional nonlinear physics problems, including the study of fluid dynamics governed by the viscous Burgers\textquotesingle{} equation, where the squared residual cost function is utilized, and the exploration of the ground state of the nonlinear Schr\"{o}dinger equation, which employs the expectation cost function. Our analysis demonstrates that a gradient-free optimization process specifically tailored to variational problems significantly outperforms existing generic optimization algorithms, such as COBYLA \cite{Powell1994, *Powell1998, *Powell2007}.

\par The rest of this article is structured as follows: Sec. \ref{Sec:Efficient_Estimation} introduces our approach for efficiently determining the cost functions and their derivatives and performing sequential optimization of the variational parameters. In particular, we systematically express the PQC as a weighted sum of distinct unitaries in Sec. \ref{Sec:PQC}, enabling the efficient construction of cost functions and their derivatives, namely the squared residual cost function discussed in Sec. \ref{Sec:Residual_Cost_Function} and the expectation cost function presented in Sec. \ref{Sec:Expectation_Cost_Function}. Subsequently, in Sec. \ref{Sec:Direct_Optimization}, we discuss an optimization protocol termed \emph{sequential grid-based explicit optimization} (SGEO), offering a simple yet powerful approach to parameter optimization in VQAs. We demonstrate our approach in two applications: the nonlinear dynamics of fluid configurations governed by the one-dimensional Burgers\textquotesingle{} equation in Sec. \ref{Sec:Burgers} and the ground state approximation of the one-dimensional nonlinear Schr\"{o}dinger equation in Sec. \ref{Sec:NLSE}. Finally, we summarize in Sec. \ref{Sec_Conclusion} and outline potential directions for future research.

\section{ Efficient Estimation and Sequential Optimization of Cost Functions }
\label{Sec:Efficient_Estimation}

\par In this section, we present a robust methodology for the efficient estimation of various cost functions and their derivatives within variational algorithms. We commence by detailing the structure of the parameterized quantum circuit (PQC) in Sec. \ref{Sec:PQC}. This is followed by the formulation of the squared residual cost function in Sec. \ref{Sec:Residual_Cost_Function} and the expectation cost function in Sec. \ref{Sec:Expectation_Cost_Function}. Finally, we outline an optimization protocol in Sec. \ref{Sec:Direct_Optimization} that we will employ in the subsequent sections for various applications.

\subsection{ Parameterized Quantum Circuits }
\label{Sec:PQC}

To present our approach, we consider a generic quantum ansatz $\hat{U}({\boldsymbol\lambda})$, parameterized by a set of variational parameters ${\boldsymbol\lambda}$, expressed as
\bea\bal \label{EQ:General_PQC}
\hat{U}({\boldsymbol\lambda}) &= \hat{\tilde{U}}_{m}(\lambda_{m}) \hat{\tilde{U}}_{m^{\prime}}^{\rm ent}\cdots \hat{\tilde{U}}_{k}(\lambda_{k}) \hat{\tilde{U}}_{j}(\lambda_{j})\cdots \hat{\tilde{U}}_{i^{\prime}}^{\rm ent} \hat{\tilde{U}}_{i}(\lambda_i)\;,~
\eal\eea
where $m$ represents the total number of parameterized single-qubit operators in the PQC. The operator $\hat{\tilde{U}}_{j}(\lambda_{j}) = e^{-i\lambda_{j}\hat{P}_{j}/2}$ represents a single-qubit Pauli rotation, where $\hat{P} \in \{\hat{X}, \hat{Y}, \hat{Z}\}$ such that $\hat{P}^{2} = \bf{1}$. This operator can also be expressed as $\hat{\tilde{U}}_{j}(\lambda_{j}) = \cos(\lambda_{j}/2)\hat{I} - i\sin(\lambda_{j}/2)\hat{P}_{j}$ \cite{Nielsen2010} or equivalently $\hat{\tilde{U}}_{j}(\lambda_{j}) = \cos(\lambda_{j}/2)\hat{\tilde{U}}_{j}(\lambda_{j} = 0) + \sin(\lambda_{j}/2)\hat{\tilde{U}}_{j}(\lambda_{j} = \pi)$, where $\hat{\tilde{U}}_{j}(\lambda_{j} = 0) \equiv \hat{I}$ and $\hat{\tilde{U}}_{j}(\lambda_{j} = \pi) \equiv -i\hat{P}_{j}$. Additionally, $\hat{\tilde{U}}_{i^{\prime}}^{\rm ent}$ denotes an unparameterized unitary that may include a combination of single-qubit operations along with two- and/or multi-qubit entangling gates. It is essential to emphasize that although we focused on parameterized single-qubit operators, a similar analysis can be performed for parameterized two-qubit gates (refer to Appendix \ref{AppSec:UOperators} for further details).

\par The PQC $\hat{U}({\boldsymbol\lambda})$ captures two scenarios. First, it may be structured as a sequence of single- and two-qubit gates tailored to the requirements of specific hardware platforms, as commonly observed in hardware-efficient ans\"{a}tze (HEA) \cite{Kandala2017}. Second, the PQC may consist of products of multi-qubit Pauli operators, typically used in problem-inspired variational Hamiltonian ans\"{a}tze (VHA) \cite{Wecker2015}; these multi-qubit unitaries can always be systematically decomposed into sequences of single-qubit rotation gates and unparameterized two-qubit entangling gates by utilizing the elements of a universal gate set \cite{Barenco1995, Nielsen2010}. These decompositions transform the VHA into the form of Eq. (\ref{EQ:General_PQC}) under specific constraints, such as $\lambda_{k} = \lambda_{j}$ or $\hat{P}_{k} \neq \hat{P}_{j}$, which arise naturally from the structure of the ansatz and constituents of the universal gate set. Thus, it is warranted to assert that Eq. (\ref{EQ:General_PQC}) represents a generic structure of a PQC. 

\par Irrespective of the internal structure of the PQC [Eq. (\ref{EQ:General_PQC})], the role of one, two, or all single-qubit parameterized operators can be analyzed, facilitating systematic formulations of the ansatz as linear combinations of distinct unitaries for various parameter configurations. These systematic formulations enhance the flexibility of subsequent discussions, allowing for a thorough exploration of various scenarios. From the perspective of a single-component $\hat{\tilde{U}}_{j}(\lambda_{j})$ in the quantum ansatz, parameterized by $\lambda_{j}$, the PQC $\hat{U}({\boldsymbol\lambda})$ can be expressed as a linear combination of two unitaries \cite{Nielsen2010} and takes the form
\bea\bal \label{EQ:General_PQC1}
\hat{U}({\boldsymbol\lambda}) = \cos(\lambda_{j}/2)\hat{U}^{0}({\boldsymbol\lambda}_{1}) + \sin(\lambda_{j}/2)\hat{U}^{\pi}({\boldsymbol\lambda}_{1}) \;,
\eal\eea
where we adopt the notation $\hat{U}^{\lambda_{j_{0}}}({\boldsymbol\lambda}_{1}) \equiv \hat{U}(\lambda_{j} = \lambda_{j_0}, {\boldsymbol\lambda}_{1})$ and $\lambda_{j_{0}} \in \{0, \pi\}$. Here, ${\boldsymbol\lambda}_{1}$ represents the set of variational parameters, temporarily fixed at their respective values, excluding $\lambda_{j}$. Eq. (\ref{EQ:General_PQC1}) provides an alternative expression of the PQC, emphasizing the contribution of a single variational parameter to its structure. From Eq. (\ref{EQ:General_PQC1}), it is easy to show that $\hat{U}({\boldsymbol\lambda})\hat{U}^{\dagger}({\boldsymbol\lambda}) = \hat{U}^{\dagger}({\boldsymbol\lambda})\hat{U}({\boldsymbol\lambda}) = \boldsymbol{1}$ and $\bigl[\hat{U}^{0}({\boldsymbol\lambda}_{1})\hat{U}^{\pi^{\dagger}}({\boldsymbol\lambda}_{1})\bigl]^{\dagger} = -\hat{U}^{0}({\boldsymbol\lambda}_{1})\hat{U}^{\pi^{\dagger}}({\boldsymbol\lambda}_{1})$. 

\par Similarly, in a setting where two single-qubit operators, $\hat{\tilde{U}}_{j}(\lambda_{j})$ and $\hat{\tilde{U}}_{k}(\lambda_{k})$, parameterized by $\lambda_{j}$ and $\lambda_{k}$, respectively, are isolated for analysis, the PQC can be written as a sum of scalar multiples of four unitaries and given as
\bea\bal \label{EQ:General_PQC2}
\hat{U}({\boldsymbol\lambda}) =&~\cos(\lambda_{j}/2)\cos(\lambda_{k}/2)\hat{U}^{0, 0}({\boldsymbol\lambda}_{2}) \\ 
& ~+  \cos(\lambda_{j}/2)\sin(\lambda_{k}/2)\hat{U}^{0, \pi}({\boldsymbol\lambda}_{2}) \\ 
& ~+ \sin(\lambda_{j}/2)\cos(\lambda_{k}/2)\hat{U}^{\pi, 0}({\boldsymbol\lambda}_{2}) \\ 
& ~+ \sin(\lambda_{j}/2)\sin(\lambda_{k}/2)\hat{U}^{\pi, \pi}({\boldsymbol\lambda}_{2}) \;.~~~~~
\eal\eea
Here, $\hat{U}^{\lambda_{j_{0}}, \lambda_{k_{0}}}({\boldsymbol\lambda}_{2}) \equiv \hat{U}(\lambda_{j} = \lambda_{j_0}, \lambda_{k} = \lambda_{k_0}, {\boldsymbol\lambda}_{2})$ and ${\boldsymbol\lambda}_{2}$ is the set of variational parameters excluding $\lambda_{j}$ and $\lambda_{k}$. It is worth noting that a similar expression arises when considering two distant single-qubit rotation operators, such as $\hat{\tilde{U}}_{j}(\lambda_{j})$ and $\hat{\tilde{U}}_{i}(\lambda_{i})$, or $\hat{\tilde{U}}_{k}(\lambda_{k})$ and $\hat{\tilde{U}}_{i}(\lambda_{i})$, which can be analyzed analogously. Furthermore, when a constraint $\lambda_{k} = \lambda_{j}$ is imposed in Eq. (\ref{EQ:General_PQC2}), as is often encountered in decomposed VHA, the expression simplifies to
\bea\bal \nonumber
\hat{U}({\boldsymbol\lambda}) &=\cos^{2}(\lambda_{j}/2)\hat{U}^{0, 0}({\boldsymbol\lambda}_{2}) + \sin^{2}(\lambda_{j}/2)\hat{U}^{\pi, \pi}({\boldsymbol\lambda}_{2})\\ & +  \cos(\lambda_{j}/2)\sin(\lambda_{j}/2)\bigl[\hat{U}^{\pi, 0}({\boldsymbol\lambda}_{2}) + \hat{U}^{0, \pi}({\boldsymbol\lambda}_{2})\bigl] \;.~~
\eal\eea
Such constraints are useful for restricting the Hilbert space of the ansatz and minimizing the optimization complexity \cite{Wecker2015}.

\par Following the same pattern and considering all $m$ parameterized single-qubit operators, the PQC can be expressed as a weighted sum of $2^{m}$ unitaries and takes the form
\bea\bal \label{EQ:General_PQC3}
\hat{U}({\boldsymbol\lambda}) = \sum_{k = 1}^{2^{m}}~\bigl[~ \prod_{l=1}^{m} \cos(\frac{\lambda_{l} - \lambda^{\prime}_{k, l}}{2})~\bigl]~U^{{\boldsymbol\lambda}^{\prime}_{k}} \;,
\eal\eea
where ${\boldsymbol\lambda}^{\prime}_{k}$ is the $k^{\rm th}$-set of fixed parameters with different permutations of values in $\{0, \pi\}$. Here, $\lambda_{l}$ ($\lambda^{\prime}_{k, l}$) is the $l^{\rm th}$ member of the set of the variational (fixed) parameters ${\boldsymbol\lambda}$ (${\boldsymbol\lambda}_{k}^{\prime}$). A notable limitation of simultaneously analyzing a large number of parameterized unitaries is the exponential increase in the number of terms in the summation relative to the number of single-qubit operators in the PQC. While constraints inspired by VHA may reduce the number of coefficients required, they do not decrease the number of unitaries in the summation. Consequently, they fail to address the inherent quantum computational complexity. In subsequent sections, we illustrate how different formulations of the ansatz $\hat{U}({\boldsymbol\lambda})$ facilitate efficient analysis of the nonlocal behavior of various commonly used cost functions and their arbitrary derivatives.

\subsection{ Case I: Squared Residual of Variational States }
\label{Sec:Residual_Cost_Function}

\par In numerous applications of VQAs, including fluid dynamics \cite{Lubasch2020, Jaksch2023}, quantum dynamics \cite{Cirstoiu2020, Lin2021, Linteau2024}, and quantum finance \cite{Sarma2024}, the cost function is typically defined as the squared residual of the variational state with respect to a specific target state. This cost function quantifies the deviation of the variational state $\vert\Psi_{\rm var}(\boldsymbol\lambda)\rangle = \hat{U}(\boldsymbol\lambda) \vert 0 \rangle$ from the desired target state $\vert\Psi_{\rm target}\rangle = \hat{U}_{\rm target} \vert 0 \rangle$, and written as
\bea\bal \label{EQ:Residual_Cost}
\mathcal{C}_{I}({\boldsymbol\lambda}) & = \vert\vert~ \vert\Psi_{\rm target}\rangle - \vert\Psi_{\rm var}(\boldsymbol\lambda)\rangle~ \vert\vert^{2} \;, \\
& = 2\bigl[~1 - {\rm Re}\{ \langle 0 \vert \hat{U}^{\dagger}_{\rm target} \hat{U}(\boldsymbol\lambda)\vert 0 \rangle \}~\bigl] \;,
\eal\eea
where ${\rm Re}\{\cdots\}$ denotes the real part of the overlap $\langle 0 \vert \hat{U}^{\dagger}_{\rm target} \hat{U}(\boldsymbol\lambda)\vert 0 \rangle$, which can be estimated using a quantum device. Conventionally, Eq. (\ref{EQ:Residual_Cost}) can be utilized to evaluate the cost function at a specific point in the high-dimensional parameter space and to compute the first- and higher-order derivatives of the cost function in the surrounding region using conventional methods such as the parameter-shift rule \cite{Schuld2019, Crooks2019, Mari2021, Banchi2021, Hubregtsen2022} and analytical gradient techniques \cite{Mitarai2019, Harrow2021}. Notably, the evaluation of the cost function at shifted parameter values using the parameter-shift rule or the computation of deep quantum circuits for analytical methods necessitates additional quantum resources. Moreover, these techniques are confined to offering insights within the immediate vicinity of a point in the parameter space and fail to characterize the broader structure of the cost function landscape.

\par First, we focus on a single-qubit operator $\hat{\tilde{U}}_{j}$ parameterized by $\lambda_{j}$, while keeping all other parameters fixed. This allows the PQC to be expressed as a linear combination of two unitaries, as outlined in Eq. (\ref{EQ:General_PQC1}). The squared residual cost function and its arbitrary derivatives with respect to parameter $\lambda_{j}$ are then given as 
\bea \label{EQ:Residual_Cost1}
\mathcal{C}_{I}({\boldsymbol\lambda}) &=~2\big[~ 1 - \cos(\lambda_{j}/2)\alpha^{0} - \sin(\lambda_{j}/2)\alpha^{\pi}~\bigl] \;, \\ \label{EQ:Residual_Cost2}
\frac{\partial^{n_{j}}{\mathcal{C}_{I}({\boldsymbol\lambda})}}{\partial{\lambda_{j}}^{n_{j}}} &= -2\bigl[~ \frac{\partial^{n_{j}}{\cos(\lambda_{j}/2)}}{\partial{\lambda_{j}}^{n_{j}}}\alpha^{0}  + \frac{\partial^{n_{j}}{\sin(\lambda_{j}/2)}}{\partial{\lambda_{j}}^{n_{j}}}\alpha^{\pi}~\bigl] \;.~~
\eea
Here, $\alpha^{\lambda_{j_0}} \equiv {\rm Re}\{ \langle 0 \vert \hat{U}^{\dagger}_{\rm target} \hat{U}^{\lambda_{j_0}}({\boldsymbol\lambda}_{1})\vert 0 \rangle \}$ can be estimated using NISQ devices, after which Eqs. (\ref{EQ:Residual_Cost1} - \ref{EQ:Residual_Cost2}) are evaluated on a classical computer within the parameter domain $\lambda_{j} \in [-\pi, \pi)$. Eq. (\ref{EQ:Residual_Cost1}) highlights that, while representing the ansatz as a linear combination of two unitaries doubles the required quantum computational resources compared to a direct evaluation of the cost function, it facilitates the estimation of the relatively nonlocal behavior of the cost function, particularly along a single dimension in the high-dimensional parameter space. Furthermore, it enables the evaluation of arbitrary $n_{j}^{\rm th}$-order derivatives of the cost function at any point along the $j^{\rm th}$-dimension of the parameter space, as described by Eq. (\ref{EQ:Residual_Cost2}), without necessitating additional quantum computational resources. In other words, Eqs. (\ref{EQ:Residual_Cost1} - \ref{EQ:Residual_Cost2}) provide significantly deeper insights into the structure of the cost function landscape and its derivatives compared to Eq. (\ref{EQ:Residual_Cost}) and conventional gradient methods \cite{Schuld2019, Crooks2019, Mari2021, Banchi2021, Hubregtsen2022, Mitarai2019, Harrow2021}, respectively. Our approach enhances efficiency in terms of the utilization of quantum resources compared to the information acquired on the cost function and its derivatives. This not only reduces resource usage but may also accelerate the practical applications of quantum technologies. Subsequently, Eqs. (\ref{EQ:Residual_Cost1} - \ref{EQ:Residual_Cost2}) can be utilized in multiple ways for both gradient-based and gradient-free classical optimization algorithms. 

\par Second, we examine the case where two arbitrary single-qubit operators, $\hat{\tilde{U}}_{j}$ and $\hat{\tilde{U}}_{k}$, parameterized by $\lambda_{j}$ and $\lambda_{k}$, are considered, while all other parameters remain fixed. Under this assumption, the unitary ansatz can be represented in the form of Eq. (\ref{EQ:General_PQC2}). The cost function and its $(n_{j}^{\rm th} + n_{k}^{\rm th})$-order derivatives then take the form 
\begin{widetext}
\bea\bal \label{EQ:Residual_Cost3}\nonumber 
\mathcal{C}_{I}({\boldsymbol\lambda}) &= 2 - 2\Bigl[~\cos(\lambda_{j}/2)\cos(\lambda_{k}/2)\beta^{0, 0} + \cos(\lambda_{j}/2)\sin(\lambda_{k}/2)\beta^{0, \pi} \\ &~~~~~~~~~~~~ + \sin(\lambda_{j}/2)\cos(\lambda_{k}/2)\beta^{\pi, 0} + \sin(\lambda_{j}/2)\sin(\lambda_{k}/2)\beta^{\pi, \pi} ~\Bigl]\;, \\
\frac{\partial^{n_{j}+n_{k}}{\mathcal{C}_{I}({\boldsymbol\lambda})}}{\partial{\lambda_{j}}^{n_{j}}\partial{\lambda_{k}}^{n_{k}}} &= -2\Bigl[~\frac{\partial^{n_{j}}\cos(\lambda_{j}/2)}{\partial{\lambda_{j}}^{n_{j}}}\frac{\partial^{n_{k}}\cos(\lambda_{k}/2)}{\partial{\lambda_{k}}^{n_{k}}}\beta^{0, 0} + \frac{\partial^{n_{j}}\cos(\lambda_{j}/2)}{\partial{\lambda_{j}}^{n_{j}}}\frac{\partial^{n_{k}}\sin(\lambda_{k}/2)}{\partial{\lambda_{k}}^{n_{k}}}\beta^{0, \pi} \\ 
&~~~~~~~~~ + \frac{\partial^{n_{j}}\sin(\lambda_{j}/2)}{\partial{\lambda_{j}}^{n_{j}}}\frac{\partial^{n_{k}}\cos(\lambda_{k}/2)}{\partial{\lambda_{k}}^{n_{k}}}\beta^{\pi, 0} + \frac{\partial^{n_{j}}\sin(\lambda_{j}/2)}{\partial{\lambda_{j}}^{n_{j}}}\frac{\partial^{n_{k}}\sin(\lambda_{k}/2)}{\partial{\lambda_{k}}^{n_{k}}}\beta^{\pi, \pi}~\Bigl]\;,
\eal\eea 
\end{widetext}
where $\beta^{\lambda_{j_{0}}, \lambda_{k_{0}}} \equiv {\rm Re}\{\langle{0}\vert{\hat{U}^{\dagger}_{\rm target}}\hat{U}^{\lambda_{j_{0}}, \lambda_{k_{0}}}({\boldsymbol\lambda}_{2})\vert{0}\rangle\}$ can be computed on NISQ devices. Analogous to the single-component case, the two-component scenario provides significant insights into the nonlocal characteristics of the cost function landscape, particularly within a two-dimensional plane of the higher-dimensional parameter space. The two-component scheme also necessitates fewer quantum resources than conventional methods to obtain the same information regarding the cost function and its derivatives. Moreover, derivatives of arbitrary order $n_{j} + n_{k}$ can be evaluated without the need for additional quantum computational resources. 

\par Lastly, we examine the scenario involving the simultaneous optimization of all parameterized single-qubit operators. This approach necessitates the estimation of an exponentially large number of $2^{m}$ terms using quantum devices. The cost function and its derivatives take the form 
\bea\bal \label{EQ:Residual_Cost4}\nonumber
\mathcal{C}_{I}({\boldsymbol\lambda}) &= 2 - 2\sum_{k=1}^{2^m}~\bigl[~\prod_{l = 1}^{m}\cos(\frac{\lambda_{l} - \lambda^{\prime}_{k, l}}{2}) ~\bigl] ~\gamma^{{\boldsymbol\lambda}^{\prime}_{k}}\;, \\
\frac{\partial^{\sum_{n_{l}}}{\mathcal{C}_{I}({\boldsymbol\lambda})}}{\prod\partial{\lambda_{l}}^{n_{l}}} &= -2\sum_{k=1}^{2^m}~\bigl[~\frac{\prod_{l = 1}^{m}\partial^{n_{l}}\cos(\frac{\lambda_{l} - \lambda^{\prime}_{k, l}}{2})}{{\prod\partial{\lambda_{l}}^{n_{l}}}}~\bigl] ~\gamma^{{\boldsymbol\lambda}^{\prime}_{k}}\;,~~
\eal\eea
where ${\boldsymbol\lambda}^{\prime}_{k}$ is the $k^{\rm th}$ set of fixed parameters where each element is either zero or $\pi$ with different permutations. Furthermore, $\lambda^{\prime}_{k, l}$ is the $l^{\rm th}$ element of $k^{\rm th}$ set of fixed parameters, $\gamma^{{\boldsymbol\lambda}^{\prime}_{k}} \equiv {\rm Re}\{ \langle{0}\vert{\hat{U}^{\dagger}_{\rm target}}\hat{U}^{{\boldsymbol\lambda}^{\prime}_{k}}\vert{0}\rangle\}$, and $\hat{U}^{{\boldsymbol\lambda}^{\prime}_{k}}$ is unitary operator with fixed parameters. It is important to emphasize that the simultaneous consideration of all parameters facilitates the exact optimization of the cost function, albeit at the expense of evaluating an exponentially large number of $2^{m}$ terms on quantum devices. In contrast, existing conventional approaches, such as gradient-based and gradient-free optimization algorithms, are hindered by issues like local minima \cite{Bittel2021, Fontana2022, Anschuetz2022} and barren plateaus \cite{Mcclean2018, Larocca2022, *Larocca2024, Qi2023} and fail to ensure convergence to the global minimum of the cost function.

\par In Sec. \ref{Sec:Direct_Optimization}, we introduce a gradient-free iterative approach called sequential grid-based explicit optimization (SGEO), which optimizes each variational parameter using Eq. (\ref{EQ:Residual_Cost1}). This iterative framework requires the evaluation of $2mR$ quantum circuits, where $R$ represents the total number of iterations across the set of variational parameters. In Sec. \ref{Sec:Burgers}, we apply this methodology to investigate the nonlinear dynamics of fluid configurations governed by the one-dimensional Burgers\textquotesingle{} equation within a variational framework.

\subsection{ Case II: Expectation Values of Observables }
\label{Sec:Expectation_Cost_Function}

\par Another widely adopted cost function in VQAs is the expectation value of an observable, which is considered for various problems across diverse domains, including quantum chemistry \cite{Peruzzo2014, Kandala2017, Wecker2015}, quantum approximate optimization algorithm (QAOA) \cite{Farhi2014}, and quadratic unconstrained binary optimization (QUBO) \cite{Tan2021}. The general form of the expectation value of an observable can be expressed as
\bea\bal \label{EQ:Expectation_Cost}
\mathcal{C}_{O}({\boldsymbol\lambda}) &= \langle\Psi_{\rm var}(\boldsymbol\lambda)\vert\hat{O}\vert\Psi_{\rm var}(\boldsymbol\lambda)\rangle = \langle{0}\vert{\hat{U}^{\dagger}(\boldsymbol\lambda)}\hat{O}{\hat{U}(\boldsymbol\lambda)}\vert{0}\rangle\;,~
\eal\eea
where $\hat{O}$ denotes a Hermitian operator, such as the Hamiltonian for a chemical system in quantum chemistry applications or an encoded system for optimization problems. Below, we express the PQC as a linear combination of unitaries and formulate the expectation cost function, along with its derivatives, tailored for various scenarios that entail the optimization of single or multiple one-qubit parameterized operators. 
\vspace{4.0mm}

\par First, we attend to the scenario of optimizing individual parameterized single-qubit unitaries and substitute the ansatz $\hat{U}(\boldsymbol\lambda)$ from Eq. (\ref{EQ:General_PQC1}) in Eq. (\ref{EQ:Expectation_Cost}). Subsequently, we derive the expectation cost function and its $n_{j}^{\rm th}$-order derivative with respect to the parameter $\lambda_{j}$, resulting in
\bea\bal \label{EQ:Expectation_Cost1}
\mathcal{C}_{O}({\boldsymbol\lambda}) =& \cos^{2}(\lambda_{j}/2)\kappa^{0, 0} + \sin^{2}(\lambda_{j}/2)\kappa^{\pi, \pi} \\ 
&+ 2\cos(\lambda_{j}/2)\sin(\lambda_{j}/2){\rm Re}\{\kappa^{0, \pi}\}\;,
\eal\eea
\bea\bal \label{EQ:Expectation_Cost2}
\frac{\partial^{n_{j}}\mathcal{C}_{O}({\boldsymbol\lambda})}{\partial{\lambda_j}^{n_{j}}} =& \frac{\partial^{n_{j}}{\cos^{2}(\lambda_{j}/2)}}{\partial{\lambda_j}^{n_{j}}}\kappa^{0, 0} + \frac{\partial^{n_{j}}{\sin^{2}(\lambda_{j}/2)}}{\partial{\lambda_j}^{n_{j}}}\kappa^{\pi, \pi} \\ 
&+ 2\frac{\partial^{n_{j}}{\cos(\lambda_{j}/2)\sin(\lambda_{j}/2)}}{\partial{\lambda_j}^{n_{j}}}{\rm Re}\{\kappa^{0, \pi}\}\;,
\eal\eea
where $\kappa^{\lambda_{j_0}^{B}, \lambda_{j_0}^{K}} \equiv \langle{0}\vert{\hat{U}^{\lambda_{j_0}^{B^{\dagger}}}({\boldsymbol\lambda}_{1})}\hat{O}{\hat{U}^{\lambda_{j_0}^{K}}({\boldsymbol\lambda}_{1})}\vert{0}\rangle$, and ${\boldsymbol\lambda}_{1}$ represents the set of variational parameters excluding $\lambda_{j}$. Here, $K$ and $B$ serve to distinguish the state and its dual in the bracket notation. It is important to note that $\kappa^{0, 0}$ and $\kappa^{\pi, \pi}$ represent expectation values and, therefore, require the execution of one quantum circuit each for their evaluation. Furthermore, in deriving Eq. (\ref{EQ:Expectation_Cost1}) and Eq. (\ref{EQ:Expectation_Cost2}), we utilized the relation $\bigl[\kappa^{0, \pi} + \kappa^{\pi, 0}\bigl] = 2{\rm Re}\{\kappa^{0, \pi}\} = 2{\rm Re}\{\kappa^{\pi, 0}\}$, which reduces computational overhead by simplifying the evaluation process from four quantum circuits (used to estimate two complex numbers) to one quantum circuit for determining a single real component. Although Eq. (\ref{EQ:Expectation_Cost1}) necessitates a three-fold increase in quantum resources compared to what is required for computing the expectation cost function as outlined in Eq. (\ref{EQ:Expectation_Cost}), it yields significantly more information regarding the behavior of the cost function landscape along the $\lambda_{j}$ dimension. Furthermore, the execution of Eq. (\ref{EQ:Expectation_Cost2}) demands no extra quantum resources for computing any derivative with respect to the parameter $\lambda_{j}$. It is crucial to highlight that the Refs. \cite{Parrish2019, Nakanishi2020, Ostaszewski2021} present a methodology where the expectation cost function and its derivatives over the entire domain of parameterized single-qubit operators are estimated. This estimation is achieved by evaluating the values of the cost function at three distinct points within the parameter domain. In contrast, our systematic formulation of the parameterized quantum circuit (PQC) provides a more comprehensive and versatile approach that not only enhances the optimization of single-qubit operators but also facilitates the extension of our method to multi-qubit gates, including the parameterized two-qubit gates discussed in Appendix \ref{AppSec:UOperators}.

\par Second, we examine the scenario where two parameterized single-qubit operators are optimized concurrently. By substituting Eq. (\ref{EQ:General_PQC2}) into Eq. (\ref{EQ:Expectation_Cost}), the expectation cost function takes the form
\begin{widetext}
\bea\bal \label{EQ:Expectation_Cost3} \nonumber
\mathcal{C}_{O}({\boldsymbol\lambda}) =&~~~ 
\cos^{2}(\lambda_{j}/2) \Bigl[ \cos^{2}(\lambda_{k}/2) \zeta^{0, 0, 0, 0} + 2\cos(\lambda_{k}/2) \sin(\lambda_{k}/2){\rm Re}\{ \zeta^{0, 0, 0, \pi} \} + \sin^{2}(\lambda_{k}/2) \zeta^{0, \pi, 0, \pi} \Bigl]\\
& + \sin^{2}(\lambda_{j}/2) \Bigl[ \sin^{2}(\lambda_{k}/2) \zeta^{\pi, \pi, \pi, \pi} +  2\cos(\lambda_{k}/2) \sin(\lambda_{k}/2){\rm Re}\{\zeta^{\pi, 0, \pi, \pi}\} + \cos^{2}(\lambda_{k}/2) \zeta^{\pi, 0, \pi, 0} \Big]\\
& + 2\cos(\lambda_{j}/2) \sin(\lambda_{j}/2) \Bigl[ \cos^{2}(\lambda_{k}/2){\rm Re}\{\zeta^{0,0,\pi, 0}\} + \sin^{2}(\lambda_{k}/2){\rm Re}\{\zeta^{0,\pi,\pi, \pi}\} \\ 
&~~~~~~~~~~~~~~~~~~~~~~~~~~~~~~~+ \cos(\lambda_{k}/2) \sin(\lambda_{k}/2) \bigl[ {\rm Re}\{\zeta^{0, 0, \pi, \pi}\} + {\rm Re}\{\zeta^{0, \pi, \pi, 0}\}
 \bigl] \Bigl],~~~
\eal\eea
\end{widetext}
where $\zeta^{\lambda^{B}_{j_{0}}, \lambda^{B}_{k_{0}}, \lambda^{K}_{j_{0}}, \lambda^{K}_{k_{0}}} \equiv \langle{0}\vert \hat{U^{\dagger}}^{\lambda^{B}_{j_{0}}, \lambda^{B}_{k_{0}}}({\boldsymbol\lambda}_{2}) \hat{O} \hat{U}^{\lambda^{K}_{j_{0}}, \lambda^{K}_{k_{0}}}({\boldsymbol\lambda}_{2}) \vert{0}\rangle$ and ($B$) $K$ assists to spot the state in the (bra) ket. Here, we have utilized the following relations: (i) $[\zeta^{0, 0, 0, \pi} + \zeta^{0, \pi, 0, 0}] = 2{\rm Re}\{\zeta^{0, 0, 0, \pi}\}$, (ii) $[\zeta^{\pi, 0, \pi, \pi} + \zeta^{\pi, \pi, \pi, 0}] = 2{\rm Re}\{\zeta^{\pi, 0, \pi, \pi}\}$, (iii) $[\zeta^{0,0,\pi, 0} + \zeta^{\pi,0,0, 0} ] = 2{\rm Re}\{\zeta^{0,0,\pi, 0}\}$, (iv) $[\zeta^{0,\pi,\pi, \pi} + \zeta^{\pi,\pi,0, \pi}] = 2{\rm Re}\{\zeta^{0,\pi,\pi, \pi}\}$, (v) $[\zeta^{0, 0, \pi, \pi} + \zeta^{\pi, \pi, 0, 0} ] = 2{\rm Re}\{\zeta^{0, 0, \pi, \pi}\}$, and (vi) $[\zeta^{0, \pi, \pi, 0} + \zeta^{\pi, 0, 0, \pi} ] = 2{\rm Re}\{\zeta^{0, \pi, \pi, 0}\}$, all of which are straightforward to prove. A mere tenfold increase in quantum computational resources allows for the estimation of the cost function and the evaluation of its arbitrary derivatives across a two-dimensional plane within a high-dimensional parameter space. 

\par Finally, we note that simultaneous optimization of all $m$ parameterized single-qubit operators requires the calculation of an exponential number of distinct terms, though fewer than $2^{2m}$, all of which can be processed in parallel across multiple quantum devices. The simultaneous consideration of all parameters facilitates the precise optimization of the expectation cost function, resulting in the attainment of the global optimum. However, the requirement to estimate exponential number of terms on quantum devices constitutes a significant limitation when $m$ is large. Overall, the methodology outlined in this section reveals that we can assess the nonlocal behavior of the expectation cost function and its arbitrary derivatives using significantly fewer quantum resources than with conventional parameter-shift rules and other analytical approaches. In Sec. \ref{Sec:NLSE}, we demonstrate that even when considering single-component scheme and iteratively optimizing all variational parameters, our approach consistently outperforms the existing COBYLA optimizer for the ground state problem of the time-independent nonlinear Schr\"{o}dinger equation.

\subsection{ Sequential Grid-Based Explicit Optimization }
\label{Sec:Direct_Optimization}

\par Building on the methodology presented in Secs. \ref{Sec:Residual_Cost_Function} and \ref{Sec:Expectation_Cost_Function} for various cost functions, we outline an optimization protocol, termed sequential grid-based explicit optimization (SGEO), which does not rely on gradients or the higher-order derivatives of the cost function. SEGO optimizes each variational parameter individually, iteratively cycling through all parameters to ascertain a set of optimal values, as illustrated in Algorithm \ref{Algo:OptimizationRoutine}.

\begin{algorithm}[hbt]
\caption{Sequential Grid-Based Explicit Optimization (SGEO)}\label{Algo:OptimizationRoutine}
\textbf{Input:} Initialize all parameters\;
\For{\rm Fixed iterations}{
    \For{\rm Number of variational parameter}{
        Estimate the necessary set of terms, $\alpha^{\lambda_{j_0}}$ or $\kappa^{\lambda_{j_0}^{B}, \lambda_{j_0}^{K}}$, using quantum devices\;
        Compute the cost in domain $\lambda_{j} \in [-\pi, \pi )$ using either Eq. (\ref{EQ:Residual_Cost1}) or Eq. (\ref{EQ:Expectation_Cost1})\; 
        Find the parameter $\lambda_{j}$ value that minimizes the cost function\;
        Update $\lambda_{j}$ parameter\;
        } 
    }
\Return{\rm Optimized variational parameters}
\end{algorithm}

\par We begin by initializing the variational parameters of the PQC and then optimize the first parameter while keeping the others unchanged. The necessary overlaps or expectation values, as required by Eq. (\ref{EQ:Residual_Cost1}) or Eq. (\ref{EQ:Expectation_Cost1}), are estimated on a quantum device. Subsequently, the cost function is evaluated for $\lambda_{j} \in [-\pi, \pi)$ on a classical computer, which facilitates the identification of the minima along $\lambda_{j}$ and the corresponding value of the parameter. The parameter $\lambda_{j}$ is then updated to a new value. This process is performed sequentially for all parameterized operators within the PQC. After optimizing the last operator, the cycle restarts and continues for a predetermined number of iterations. 

\par Although additional conditions could be introduced to terminate the iterations on the fly or gradient-based methods can be employed to refine the protocol, we adopt this straightforward approach for classical optimization in this work. The effectiveness of the sequential grid-based explicit optimization (SGEO) protocol is assessed by comparing its performance with that of the more advanced COBYLA optimizer \cite{Powell1994, *Powell1998, *Powell2007} in Sec. \ref{Sec:Burgers} and Sec. \ref{Sec:NLSE}.

\subsection{ Quantum Ansatz Structure }
\label{Sec:Ansatz}

\begin{figure}[b]\begin{center}
\includegraphics[clip, trim=0.5cm 0.5cm 0.5cm 1.5cm, width=1.00\linewidth, height=0.67\linewidth, angle=0]{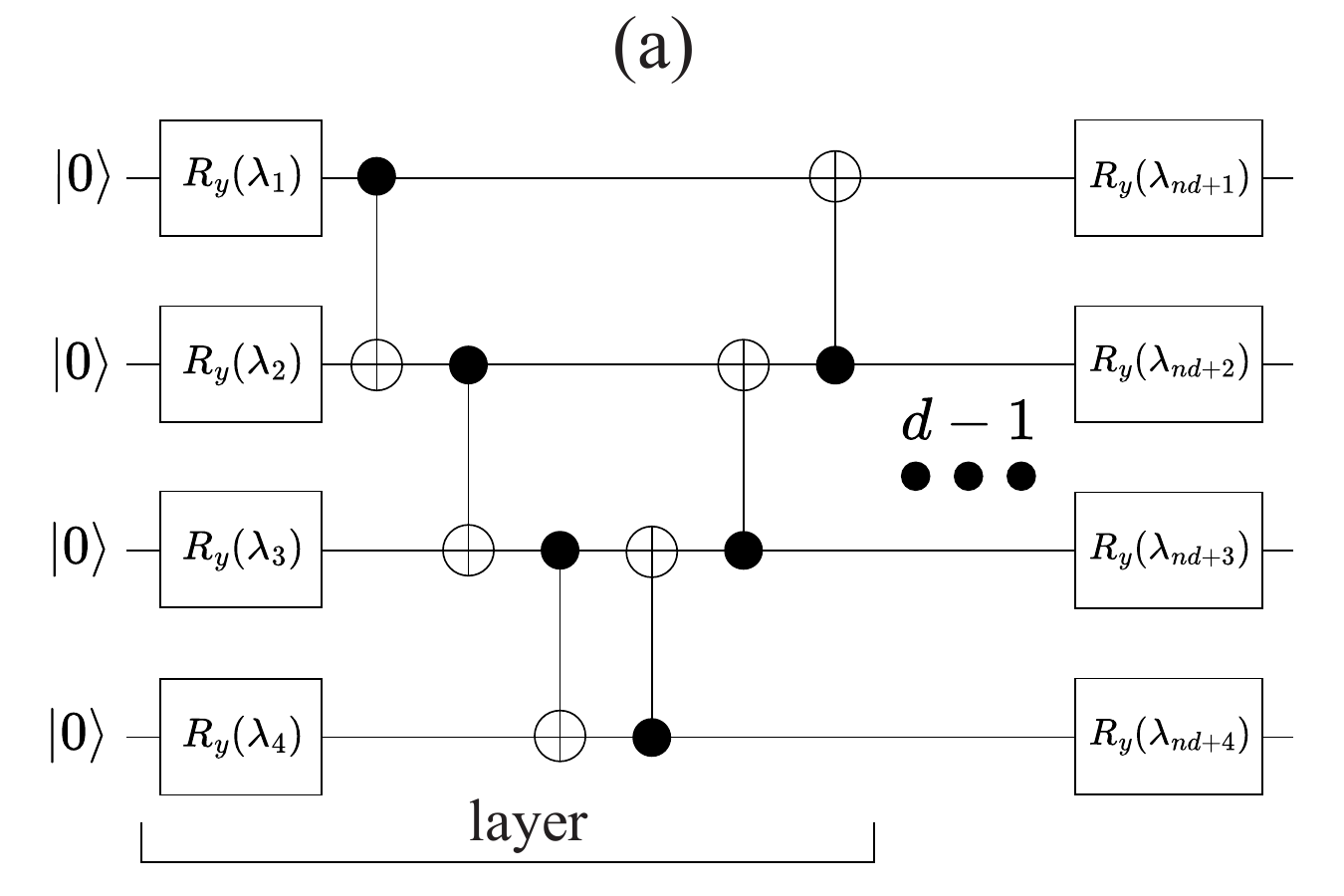}
\caption{Structure of the real-amplitude ansatz. Here, $d$ is the total number of layers and $R_{y}(\lambda_{i})$ is the single-qubit rotation gate. \vspace{-0.2cm}}
\label{Fig:HEA_Ansatz}
\end{center}\end{figure}

\par Before delving into various applications, we first introduce the real-amplitude ansatz, illustrated in Fig. \ref{Fig:HEA_Ansatz}, which will be employed throughout the remainder of this article. Here, each layer consists of $R_{y}(\lambda_{i})$ rotation gates applied uniformly across all qubits, followed by a sequence of CNOT gates. Subsequent to the final layer, $R_{y}(\lambda_{i})$ rotation gates are applied to each qubit once more, resulting in a total of $n(d+1)$ variational parameters, where $n$ denotes the number of qubits that encode the problem and $d$ the number of layers. In addition to this, we utilize the QISKIT $1.0$ platform \cite{Anis2021}, $5\times{10^{4}}$ shots per circuit, BasicSimulator for cost function evaluation, and statevector simulator to estimate the state. Throughout this article, COBYLA optimizer is utilized with the parameters ${\rm rhobeg} = \pi/16$ and ${\rm tol} = 10^{-10}$. Furthermore, it is important to note that our focus is not on assessing the expressivity of the ansatz; rather, we employ a predetermined number of layers to adequately capture the solution to the problem at hand.

\section{ Nonlinear Dynamics: The Burgers\textquotesingle{} Equation } 
\label{Sec:Burgers}

\par In this section, we focus on probing the nonlinear dynamics of fluid configurations in quantum processors using our proposed optimization framework, which is generally applicable to a range of problems, including quantum dynamics \cite{Cirstoiu2020, Lin2021, Linteau2024} and quantum finance \cite{Sarma2024}. Specifically, the nonlinear dynamics is governed by the time-dependent one-dimensional viscous Burgers\textquotesingle{} equation \cite{Bec2007, Lubasch2020, Jaksch2023}
\bea\bal \label{EQ:Burgers}
\partial_{t}u(x, t) = \bigl[\nu\partial_{x}^{2} - u(x, t)\partial_{x}\bigl]u(x, t)\;. 
\eal\eea
Here, the first term on the right side of the equation represents diffusion, where $\nu$ denotes the coefficient of kinematic viscosity. The second term corresponds to convection, which is inherently nonlinear in nature. $\partial_{x}$ and $\partial_{t}$ denote the spatial and temporal derivatives, respectively, while $u(x, t)$ is the real-valued fluid field, and $x$ ($t$) represents space (time). Eq. (\ref{EQ:Burgers}) models various flow regimes, including laminar and turbulent behavior, and small instances of the problem can be solved on a classical computer using a range of straightforward techniques, such as direct numerical simulation, finite difference methods, finite volume methods, spectral methods and other approaches. Building upon the methodology presented in Refs. \cite{Lubasch2020, Jaksch2023} in conjugation with our approach proposed in Sec. \ref{Sec:Residual_Cost_Function}, we now formulate the cost function for the variational algorithm applied to the nonlinear dynamics of fluid fields.

\par Following standard numerical approaches, we discretize the spatial (temporal) interval $[a, b]$ ($[0, T]$) into $N_{x}$ ($N_{t}$) equidistant grid points, such that $x_{n_{x}} = a + \delta_{x}{n_{x}}$ and $t_{n_{t}} = \tau{n_{t}}$. Here, $\delta_{x} = (b - a)/N_{x}$ and $\tau = T/N_{t}$ are the spacing between two adjacent spatial and temporal grid points, where $n_{i} \in \{0, 1, 2, \cdots, N_{i} - 1\}$. In this work, we impose periodic boundary conditions in spatial dimension, where $u_{0, t} = u_{N_{x}, t}$, with the values $u_{n_{x}, t}$ forming a vector $\vert{\bf u}_{t}\rangle$. Since Burgers\textquotesingle{} equation does not conserve the norm of vector $\vert{\boldsymbol{u}}_{t}\rangle$, we introduce a hyper-parameter $\Lambda_{t}$ such that $\vert{\bf u}_{t}\rangle = \Lambda_{t}\vert\Psi_{t}\rangle$, where $\vert\Psi_{t}\rangle$ is the normalized state, i.e., $\langle\Psi_{t}\vert\Psi_{t}\rangle = \langle{\bf u}_{t}\vert{\bf u}_{t}\rangle/\vert\Lambda_{t}\vert^{2} = 1$. Furthermore, spatial and temporal derivatives are approximated using the finite difference and the Euler methods, respectively, such that Eq. (\ref{EQ:Burgers}) takes the form
\bea\bal \label{EQ:Burgers1} \nonumber
\Lambda_{t+\tau}&\vert\Psi_{t+\tau}({\boldsymbol\lambda})\rangle = \\
&~~~\Bigl[\Lambda_{t} + l_{1}(\hat{A} + \hat{A}^{\dagger} - 2\hat{I}) - l_{2}\hat{D}_{t}(\hat{A} - \hat{A}^{\dagger})\Bigl]\vert\Psi_{t}\rangle\;, 
\eal\eea
where $\partial_{x}^{2} = \frac{\hat{A} + \hat{A}^{\dagger} - 2\hat{I}}{2\delta_{x}^{2}}$, $\partial_{x} = \frac{\hat{A} - \hat{A}^{\dagger}}{2\delta_{x}}$, $l_{1} = {\Lambda_{t}\tau\nu}/{2\delta_{x}^{2}}$, $l_{2} = {\vert\Lambda_{t}\vert^{2}\tau}/{2\delta_{x}}$, and $\hat{D}_{t} = {\rm diag}(\psi_{n_{x}, t})$. Here, $\hat{A}$ represents the adder operator which shifts the basis states by one unit \cite{Lubasch2020, Jaksch2023}, thereby implementing the spatial differentiation in the finite difference approximation. The squared residual cost function is then given as
\bea\bal\label{EQ:Burgers2}
C_{I}({\boldsymbol\lambda}) &= - \Bigl[\bigl(\Lambda^{*}_{t} - 2l_{1}^{*}\bigl)~{\rm Re}\{\langle{0}\vert \hat{U}_{t}^{\dagger}\hat{U}_{t + \tau}\vert{0}({\boldsymbol\lambda})\rangle\} \\ 
& ~~~~~~~+ l_{1}^{*}{\rm Re}\{\langle{0}\vert \hat{U}_{t}^{\dagger}\bigl(\hat{A} + \hat{A}^{\dagger}\bigl)\hat{U}_{t + \tau}({\boldsymbol\lambda})\vert{0}\rangle\} \\
& ~~~~~~~+ l_{2}{\rm Re}\{\langle{0}\vert \hat{U}_{t}^{\dagger}\bigl(\hat{A} - \hat{A}^{\dagger}\bigl){D_{t}^{\dagger}}\hat{U}_{t + \tau}({\boldsymbol\lambda})\vert{0}\rangle\} \Bigl]^{2}\;,~~~~
\eal\eea
where $\vert\Psi_{t+\tau}({\boldsymbol\lambda})\rangle = \hat{U}_{t+\tau}({\boldsymbol\lambda})\vert{0}\rangle$, $\vert\Psi_{t}\rangle = \hat{U}_{t}\vert{0}\rangle$, and the hyper-parameter $\Lambda_{t+\tau}$ is eliminated through optimization (see Appendix \ref{AppSec:Burger} for details). Conventionally, Eq. (\ref{EQ:Burgers2}) serves as the cost function, where one can employ gradient-based \cite{Kingma2014, Broyden1970, *Fletcher1970, *Goldfarb1970, *Shanno1970, Kraft1988} or gradient-free \cite{Powell1994, *Powell1998, *Powell2007, Spall1987, *Spall1992, Nelder1965} algorithms for optimization. The dynamics over an extended period is then analyzed by sequentially optimizing the cost function at each time step.

\par Following the single-component scheme proposed in Sec. \ref{Sec:Residual_Cost_Function}, we substitute $\hat{U}_{t + \tau}({\boldsymbol\lambda})$ in the form of Eq. (\ref{EQ:General_PQC1}), resulting in the cost function expression
\bea\bal\label{EQ:Burgers3}
C_{I}({\boldsymbol\lambda}) &= -\Bigl[\cos(\lambda_{j}/2)\bigl[G_{1}^{0} + G_{2}^{0} + G_{3}^{0}\bigl]\\
&~~~~~~~ + \sin(\lambda_{j}/2)\bigl[G_{1}^{\pi} + G_{2}^{\pi} + G_{3}^{\pi}\bigl] \Bigl]^{2}, ~~~~ 
\eal\eea
where $G^{\lambda_{j_{0}}}_{1} = \bigl(\Lambda^{*}_{t} - 2l_{1}^{*}\bigl){\rm Re}\{\langle{0}\vert \hat{U}_{t}^{\dagger}\hat{U}_{t + \tau}^{\lambda_{j_{0}}}({\boldsymbol\lambda}_{1})\vert{0}\rangle\}$, $G^{\lambda_{j_{0}}}_{2} = l_{1}^{*}{\rm Re}\{\langle{0}\vert \hat{U}_{t}^{\dagger}\bigl(\hat{A} + \hat{A}^{\dagger}\bigl)\hat{U}_{t + \tau}^{\lambda_{j_{0}}}({\boldsymbol\lambda}_{1})\vert{0}\rangle\}$, and $G^{\lambda_{j_{0}}}_{3} = l_{2}{\rm Re}\{\langle{0}\vert \hat{U}_{t}^{\dagger}\bigl(\hat{A} - \hat{A}^{\dagger}\bigl) \hat{D}^{\dagger}_{t}\hat{U}_{t + \tau}^{\lambda_{j_{0}}}({\boldsymbol\lambda}_{1})\vert{0}\rangle\}$. Here, $G_{i}^{\lambda_{j_{0}}}$\textquotesingle{s} are estimated on quantum devices, and then Eq. (\ref{EQ:Burgers3}) is evaluated on a classical computer. Eq. (\ref{EQ:Burgers3}) enables the estimation of cost function values along a line in the parameter space, specifically $\lambda_{j} \in [-\pi, \pi)$, at the expense of requiring twice the quantum resources needed to evaluate the cost function at a single point using Eq. (\ref{EQ:Burgers2}). Moreover, first-order derivative is expressed as 
\bea\bal \label{EQ:Burgers4}\nonumber
\frac{\partial{C_{I}}({\boldsymbol\lambda})}{\partial{\lambda_{j}}} &= -2\sqrt{- C_{I}({\boldsymbol\lambda})}\Bigl[~\frac{\partial\cos(\lambda_{j}/2)}{\partial{\lambda_{j}}}\bigl[~G_{1}^{0} + G_{2}^{0} + G_{3}^{0}~\bigl]\\
&~~~~~~~ + \frac{\partial\sin(\lambda_{j}/2)}{\partial{\lambda_{j}}}\bigl[~G_{1}^{\pi} + G_{2}^{\pi} + G_{3}^{\pi}~\bigl]~\Bigl], ~~~~
\eal\eea 
and it does not require any additional quantum resources. In fact, any higher-order derivative can be calculated on a classical computer without requiring more use of quantum devices. 

\begin{figure}[hbt]\begin{center}
\includegraphics[clip, trim=0.0cm 0.00cm 0.0cm 0.0cm, width=1.00\linewidth, height=1.45\linewidth, angle=0]{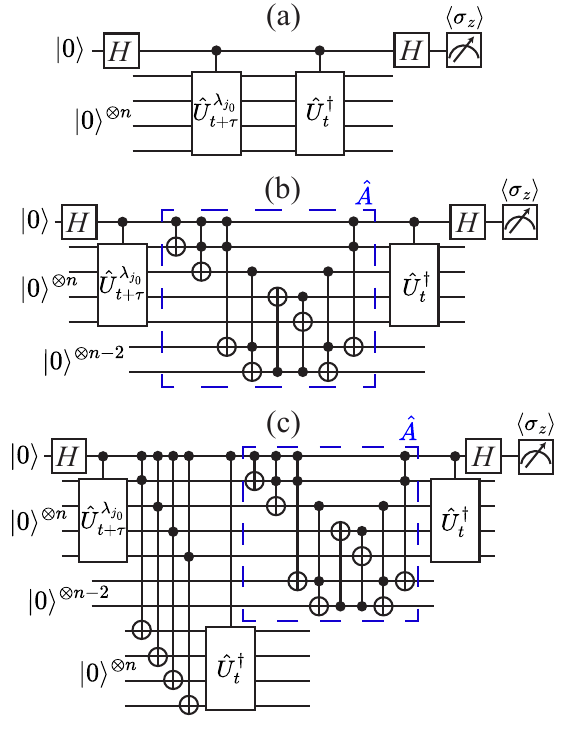}
\caption{Design of quantum circuits for evaluating various components of the cost function of the Burgers\textquotesingle{} equation. The quantum circuits in panels (a-c) measure the cost function constituents ${\rm Re}\{\langle{0}\vert \hat{U}_{t}^{\dagger}\hat{U}^{\lambda_{j_0}}_{t + \tau}\vert{0}\rangle\}$, ${\rm Re}\{\langle{0}\vert \hat{U}_{t}^{\dagger}\hat{A}\hat{U}^{\lambda_{j_0}}_{t + \tau}\vert{0}\rangle\}$, and ${\rm Re}\{\langle{0}\vert \hat{U}_{t}^{\dagger}\hat{A}{D_{t}^{\dagger}}\hat{U}^{\lambda_{j_0}}_{t + \tau}\vert{0}\rangle\}$, respectively. To evaluate ${\rm Re}\{\langle{0}\vert \hat{U}_{t}^{\dagger}\hat{A}^{\dagger}\hat{U}^{\lambda_{j_0}}_{t + \tau}\vert{0}\rangle\}$, and ${\rm Re}\{\langle{0}\vert \hat{U}_{t}^{\dagger}\hat{A}^{\dagger}{D_{t}^{\dagger}}\hat{U}^{\lambda_{j_0}}_{t + \tau}\vert{0}\rangle\}$, $\hat{A}$ in panel (b) and (c) is inverted to implement $\hat{A}^{\dagger}$. $H$ is the Hadamard gate, $\hat{A}$ is the adder circuit, and $\hat{U}^{\lambda_{j_0}}_{t+\tau} = \hat{U}^{\lambda_{j_0}}_{t+\tau}({\boldsymbol\lambda}_{1})$ is the unitary ansatz operator with $\lambda_{j} = \lambda_{j_{0}}$, while all other variational parameters remain unchanged. Here, we have demonstrated the case where $n = 4$. \vspace{-0.2cm}}
\label{Fig:QNPU_Burgers}
\end{center}\end{figure}

\par Before discussing the results of our sequential grid-based explicit
optimization (SGEO) protocol, we outline the quantum circuits employed to evaluate the $G_{i}^{\lambda_{j_{0}}}$\textquotesingle{s} terms. First, the $G_{1}^{\lambda_{j_0}}$ term (up to a constant) in Eq. (\ref{EQ:Burgers3}) represents the overlap of two quantum states, which can be measured using the quantum circuit illustrated in Fig. \ref{Fig:QNPU_Burgers}a. Here, $\hat{U}_{t}$ represents the unitary operator at time $t$, and $\hat{U}^{\lambda{j_{0}}}_{t+\tau}$ denotes the ansatz at the time instance $t+\tau$, evaluated at $\lambda_{j} = \lambda_{j_{0}}$. The component $G_{2}^{\lambda_{j_0}}$ ($G_{3}^{\lambda_{j_0}}$) comprises two terms that differ only in the operators $\hat{A}$ and $\hat{A}^{\dagger}$. For terms in the $G_{2}^{\lambda_{j_0}}$ ($G_{3}^{\lambda_{j_0}}$), we utilize the quantum circuit depicted in Fig. \ref{Fig:QNPU_Burgers}b (Fig. \ref{Fig:QNPU_Burgers}c). In these circuits, the $\hat{A}^{\dagger}$ operation is implemented by inverting the $\hat{A}$ circuit, which is enclosed in a blue dashed line in Figs. \ref{Fig:QNPU_Burgers}b-\ref{Fig:QNPU_Burgers}c. 

\par We now numerically solve the dynamics of fluid configurations for various parameter values and initial conditions. The VQA is implemented using the cost function defined in Eq. (\ref{EQ:Burgers2}) for the COBYLA optimizer and Eq. (\ref{EQ:Burgers3}) for our SGEO protocol. Additionally, we compute the infidelity $F'(t) = 1 - \vert\langle\Psi_{\rm classical}(t)\vert\Psi_{\rm opt}(t)\rangle\vert^{2}$ to quantify the deviation of the optimized variational state from the target state obtained via classical method for each time instance. While the cost function of VQA is the squared residual of the variational state defined in Eq. (\ref{EQ:Burgers2}) and Eq. (\ref{EQ:Burgers3}), throughout our analysis we rather report the infidelity $F'(t)$ of the optimized variational state $\vert\Psi_{\rm opt}(t)\rangle$ with respect to the target classical state $\vert\Psi_{\rm classical}(t)\rangle$. In this section, we adopt a time step of $\tau = \delta_{x}/10$. For the laminar regime, the optimizer iterations are chosen as $N_{\rm COBYLA} = 100$, and $N_{\rm SGEO} = 5$, while for the turbulent regime, $N_{\rm COBYLA} = 200$, and $N_{\rm SGEO} = 10$. However, it is important to note that the optimization process may terminate before all iterations are fully consumed. The iteration counts translate to $N_{\rm COBYLA}$ parameter updates for the COBYLA optimizer and $mN_{\rm SGEO}$ for the SGEO protocol, where $m$ denotes the total number of variational parameters. Finally, the number of ansatz layers is configured as $d = 2$ ($d = 3$) for $n = 3$ ($n = 4$) in the laminar regime, and $d = 3$ ($d = 4$) for $n = 3$ ($n = 4$) in the turbulent regime.

\begin{figure*}[bht]\begin{center}
\includegraphics[clip, trim=0.0cm 0.5cm 0.0cm 0.0cm, width=0.85\linewidth, height=0.75\linewidth, angle=0]{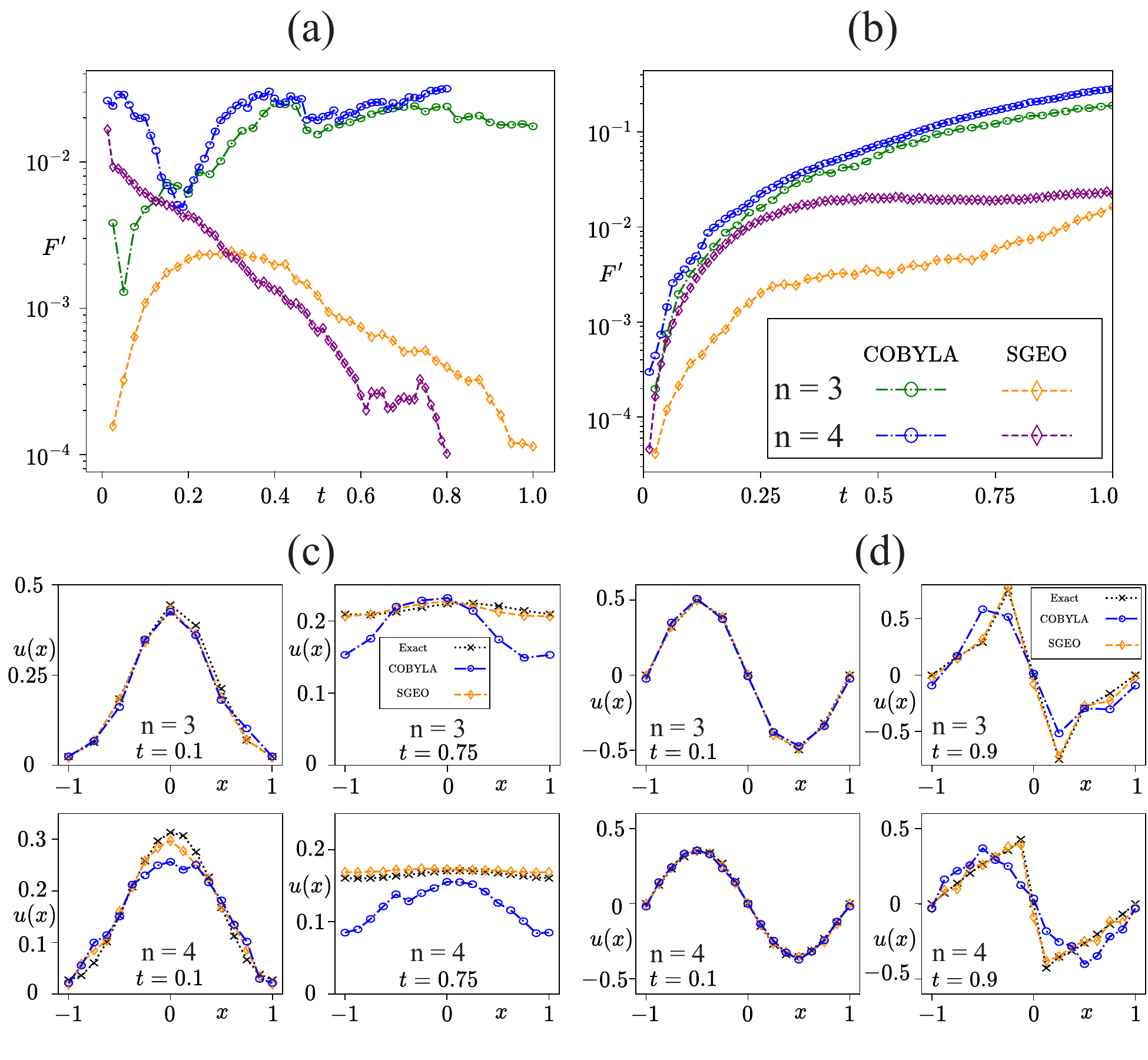}
\caption{Nonlinear dynamics of fluid configurations governed by the Burgers\textquotesingle{} equation. Panel a (b) illustrates the infidelity $F'(t) = 1 - \vert\langle\Psi_{\rm classical}(t)\vert\Psi_{\rm opt}(t)\rangle\vert^{2}$, characterizing evolved states within laminar (turbulent) regimes with kinematic viscosity $\nu = 1$ ($\nu = 10^{-3}$). This metric, $F'(t)$, serves to quantify the discrepancy between states derived from classical simulations and those produced via VQAs using COBYLA optimizer (blue and green circles) and our SGEO routine (purple and orange diamonds). Panel c (d) showcase the fluid configurations for laminar (turbulent) regimes for assorted time instances. Here, black, blue, and orange colors represent outcomes from classical method, COBYLA optimizer, and our SGEO routine, respectively. \vspace{-0.2cm}}
\label{Fig:Burgers}
\end{center}\end{figure*}

\par First, the kinematic viscosity is set to $\nu = 1$, which corresponds to the laminar regime. A square wave profile is chosen as the initial configuration, defined as $u(x, t = 0) = 1$ for $\vert{x}\vert < 0.5$ and $u(x, t = 0) = 0$ otherwise. We solve the dynamics of the Burgers\textquotesingle{} equation and present the behavior of the infidelity $F'(t)$ in Fig. \ref{Fig:Burgers}a. The results indicate that the fidelity of the variational state remains consistently above $99\%$ when utilizing our SGEO protocol and progressively increases, surpassing $99.99\%$ toward the end of the dynamics. In comparison, the COBYLA optimizer yields an initial fidelity in the range of $98\%-99.5\%$, which either briefly increases before declining, as observed for the $4$-qubit system, or decreases consistently throughout the dynamics, as seen in the $3$-qubit case. Fig. \ref{Fig:Burgers}a illustrates that our SGEO protocol consistently outperforms COBYLA optimizer at each time step. The inefficient performance of the COBYLA optimizer evidently results in gradual deviations from the target classical state at each time step, ultimately leading to the preparation of a disparate state by the end of the dynamics. We examine the evolved state by plotting the fluid configuration profiles at different time steps, as illustrated in Fig. \ref{Fig:Burgers}c. The VQA results derived from our SGEO protocol (orange diamonds) exhibit a high degree of overlap with the target classical results (black crosses). In comparison, the results obtained using the COBYLA optimizer (blue circles) produce a state that significantly diverges from the exact solution.

\par Now, we set the kinematic viscosity to $\nu = 10^{-3}$, placing the system parameters in the turbulent regime of fluid flow. The initial condition is defined as a sine wave, $u(x, t = 0)= -\sin(\pi{x})$, which evolves into a shock wave over time, as we shall observe. Fig. \ref{Fig:Burgers}b demonstrates the results of the infidelity $F'(t)$ over the course of the dynamics. As time progresses, the fidelity of the variational state declines due to the accumulation of optimization errors. This decline arises from the inherent limitations in the precision of the cost function, which is affected by the statistical shot noise. Consequently, as these errors accumulate over time, there is a certain level of deterioration in the fidelity of the variational state. This decline is particularly pronounced when employing the COBYLA optimizer. Toward the end of the dynamics at $t = 1.0$, the fidelity decreases to $98\%-99\%$ for the results obtained using our SGEO protocol, compared to a more substantial drop to $80\%$ for the results obtained utilizing the COBYLA optimizer. Fig. \ref{Fig:Burgers}d illustrates the evolved state profiles at $t = 0.1$ and $t = 0.9$. The VQA employing the SGEO protocol (orange diamonds) yields results that align closely with the classical predictions (black crosses). Moreover,  sawtooth shockwave patterns \cite{Bendaas2018} (black crosses and orange diamonds) are observed at $t = 0.9$. In contrast, the variational state obtained from the VQA using the COBYLA optimizer (blue circles) deviates significantly from the target classical state, resulting in a distinctly different state at $t = 0.9$.

\begin{figure}[htb]\begin{center}
\includegraphics[clip, trim=0.0cm 3.3cm 0.0cm 0.0cm, width=1.00\linewidth, height=0.60\linewidth, angle=0]{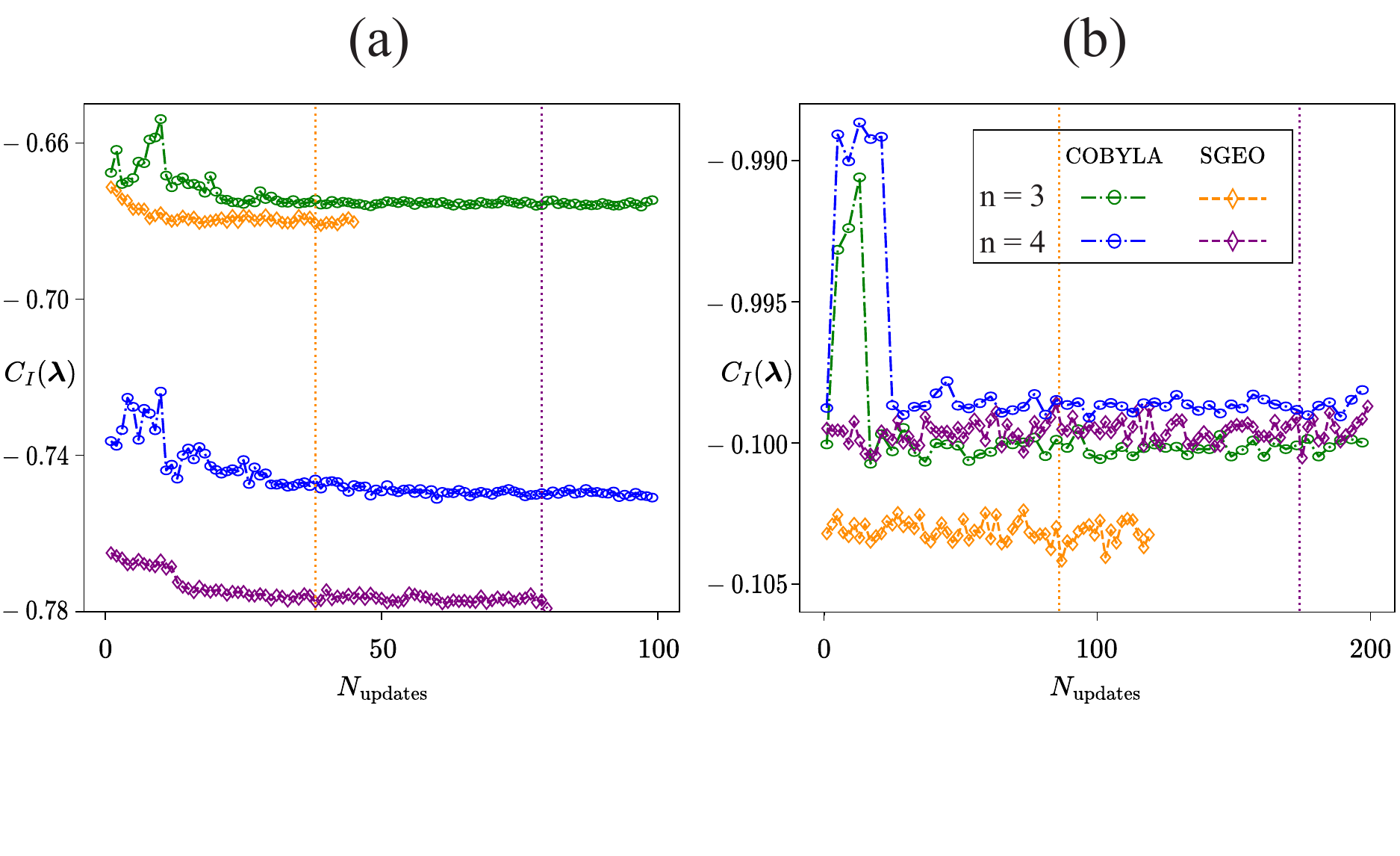}
\caption{Behavior of Burgers\textquotesingle{} equation cost function against the parameter updates. Panel (a) is for the case of laminar regime, where we took the snapshot at $t = \tau$, whereas the panel (b) is the turbulent case where we took the time instance $t = 2\tau$. The vertical lines indicate the minimum value of the cost function.  \vspace{-0.2cm}}
\label{Fig:Burgers_Cost}
\end{center}\end{figure}

\par The findings discussed above demonstrate that our SGEO protocol consistently outperforms the COBYLA optimizer across various parameter regimes. To further validate SGEO\textquotesingle{s} efficiency, Fig. \ref{Fig:Burgers_Cost} illustrates the behavior of the cost function against the parameter updates. SGEO protocol consistently achieves lower cost function values compared to COBYLA throughout the optimization process. Conversely, the COBYLA optimizer fails to attain the minimum cost function value. The superior performance of our SGEO protocol likely stems from its ability to exploit the nonlocal behavior of the cost function. This capability facilitates efficient updates to the variational parameters, enabling the generation of states that closely approximate the target classical states. The fluctuations in the cost function values, in Fig. \ref{Fig:Burgers_Cost}, are due to statistical shot noise. Furthermore, vertical lines in Fig. \ref{Fig:Burgers_Cost} indicate the minimum cost function value achieved by SGEO, illustrating that the SGEO protocol can terminate before exhausting all iterations. Future research could focus on developing a mechanism to terminate the SGEO protocol upon reaching the cost function minimum. Overall, our findings establish the SGEO protocol as a robust and reliable optimization approach for addressing complex variational problems in quantum algorithms.

\section{ Ground State Problem: The Nonlinear Schr\"{o}dinger Equation } 
\label{Sec:NLSE}

\par In this section, we demonstrate the application of our devised strategy to VQE cases involving the one-dimensional, time-independent nonlinear Schr\"{o}dinger equation, while noting that this generic approach can also be extended to other scenarios, including those in quantum chemistry \cite{Peruzzo2014, Kandala2017, Wecker2015}, QAOA \cite{Farhi2014}, and QUBO \cite{Tan2021}. The nonlinear Schr\"{o}dinger equation (NLSE) and its various extensions are used to describe a wide range of physical phenomena \cite{Agrawal2000, Nakkeeran2002, Triki2019, Sulem1999, Gross1961, Pitaevskii1961, Dalfovo1999, Leggett2001, Pitaevskii2003}, including the propagation of light in nonlinear optical systems \cite{Nakkeeran2002, Triki2019}, the formation of envelope solitons and modulation instabilities in plasma physics and surface gravity waves \cite{Sulem1999}, as well as properties such as superfluidity and vortex generation in Bose-Einstein Condensates (BEC) \cite{Gross1961, Pitaevskii1961, Dalfovo1999, Leggett2001, Pitaevskii2003}, among others. In its dimensionless form, the time-independent NLSE is expressed as 
\bea\bal \label{EQ:NLSE}
\big[-\frac{1}{2}\frac{d^{2}}{d{x}^{2}} + V(x) + g{I_{f(x)}} \big] {f(x)} = E {f(x)}\;.
\eal\eea 
Here, $f(x)$ is a normalized, real-valued function defined over the spatial domain $x \in [a, b]$. The term $I_{f(x)}$ accounts for the nonlinear interaction, $g$ specifies the nonlinearity strength, and $V(x)$ denotes the external potential. Specifically, we assume $I_{f(x)} \equiv \vert{f(x)}\vert^{2}$, with a quadratic potential $V(x) \equiv V_{0}(x - x_{0})^{2}$ centered at $x_{0} = \frac{b-a}{2}$. Additionally, periodic boundary conditions are applied, ensuring that $f(b) = f(a)$ and $V(b) = V(a)$. Numerical solutions to small instances of Eq. (\ref{EQ:NLSE}) can be obtained on classical computers through methods such as imaginary-time evolution \cite{Edwards1995, Dalfovo1999, Lubasch2018}, spectral techniques, variational approaches, or other established methods. 

\par To derive the energy cost function for ground state problem of the NLSE, we follow the procedure outlined in the previous section (also see Refs. \cite{Lubasch2020, Umer2024}). The energy cost function is expressed as the sum of the expectation values of potential, interaction, and kinetic energies, $\langle\langle{E}\rangle\rangle = \langle\langle{E_{P}}\rangle\rangle + \langle\langle{E_{I}}\rangle\rangle + \langle\langle{E_{K}}\rangle\rangle$, where
\bea\bal \label{EQ:NLSE1}
\langle\langle{E_{P}}\rangle\rangle &= \mathcal{N}~{\rm Re}\{\langle{0}\vert{\hat{U}^{\dagger}({\boldsymbol\lambda})}\hat{V}{\hat{U}({\boldsymbol\lambda})}\vert{0}\rangle\}\;,\\ 
\langle\langle{E_{I}}\rangle\rangle &= g~{\rm Re}\{\langle{0}\vert {\hat{U}^{\dagger}({\boldsymbol\lambda})} \hat{D}^{\dagger}_{2}({\boldsymbol\lambda})\hat{D}_{1}({\boldsymbol\lambda}){\hat{U}({\boldsymbol\lambda})} \vert{0}\rangle\}\bigl/\delta_{x}\;,~~~~~\\
\langle\langle{E_{K}}\rangle\rangle &= -{\rm Re}\{\langle{0}\vert{\hat{U}^{\dagger}({\boldsymbol\lambda})}\bigl[\hat{A} + \hat{A}^{\dagger} - 2\hat{I} \bigl]{\hat{U}({\boldsymbol\lambda})}\vert{0}\rangle\}\bigl/2\delta_{x}^{2}\;.~~~~~
\eal\eea
Here, $\mathcal{N} = \vert\vert{V(x)}\vert\vert$, and $\hat{D}_{1}({\boldsymbol\lambda})$ and $\hat{D}_{2}({\boldsymbol\lambda})$ are diagonal unitary operators, each encoding a copy of the variational state \cite{Lubasch2020, Umer2024}. In the variational algorithms, the total energy is used as the cost function, $\mathcal{C}_{O} = \sum_{j} \langle\langle E_{j} \rangle\rangle$, where the minimum of this cost function corresponds to the ground state energy solution. 

\par We consider single-parameter formulation of unitary ansatz $\hat{U}({\boldsymbol\lambda})$ from Eq. (\ref{EQ:General_PQC1}). The components of the energy cost function take the form 
\bea\bal \label{EQ:NLSE2}
\langle\langle{E_{P}}\rangle\rangle &=  \cos^{2}(\lambda_{j}/2)\mathcal{N}~\Gamma_{P}^{0, 0} + \sin^{2}(\lambda_{j}/2)\mathcal{N}~\Gamma_{P}^{\pi, \pi} ~~~~\\ &~~~~~+ 2\cos(\lambda_{j}/2)\sin(\lambda_{j}/2)\mathcal{N}~\Gamma_{P}^{\pi, 0} \;,~~~~
\eal\eea \vspace{-3.0mm}
\bea\bal \label{EQ:NLSE3}
\langle\langle{E_{I}}\rangle\rangle &= \frac{g}{\delta_{x}}\bigl[~\cos^{4}(\lambda_{j}/2)\Gamma_{I}^{0, 0, 0, 0} + \sin^{4}(\lambda_{j}/2)\Gamma_{I}^{\pi, \pi, \pi, \pi}\\ &~~~~~~~~~~ + 4\cos^{3}(\lambda_{j}/2)\sin(\lambda_{j}/2)\Gamma_{I}^{0, \pi, 0, 0} \\ 
&~~~~~~~~~~+ 4\sin^{3}(\lambda_{j}/2)\cos(\lambda_{j}/2)\Gamma_{I}^{\pi, 0, \pi, \pi} \\ 
&~~~~~~~~~~ + 6\cos^{2}(\lambda_{j}/2)\sin^{2}(\lambda_{j}/2)\Gamma_{I}^{0, \pi, \pi, 0}~\bigl] \;,~~~~~~
\eal\eea \vspace{-3.0mm}
\bea\bal \label{EQ:NLSE4}
\langle\langle{E_{K}}\rangle\rangle &= \frac{1}{2\delta_{x}^{2}}\bigl[~ 2 - \cos^{2}(\lambda_{j}/2)\Gamma_{K}^{0, 0} - \sin^{2}(\lambda_{j}/2)\Gamma_{K}^{\pi, \pi}~~ \\ &~~~~~~~~~~~- 2\cos(\lambda_{j}/2)\sin(\lambda_{j}/2)\Gamma_{K}^{\pi, 0}~\bigl] \;,~~
\eal\eea
where $\Gamma_{P}^{\epsilon, \eta} \equiv {\rm Re}\{\langle{0}\vert{\hat{U}^{\epsilon^{\dagger}}}\hat{V}{\hat{U}^{\eta}}\vert{0}\rangle\}$, $\Gamma_{I}^{\epsilon, \eta_{2}, \eta_{1}, \eta} \equiv {\rm Re}\{\langle{0}\vert{\hat{U}^{\epsilon^{\dagger}}}\hat{D}_{2}^{\eta_{2}^{\dagger}}\hat{D}_{1}^{\eta_{1}}{\hat{U}^{\eta}}\vert{0}\rangle\}$, and $\Gamma_{K}^{\epsilon, \eta} \equiv {\rm Re}\{\langle{0}\vert{\hat{U}^{\epsilon^{\dagger}}}\bigl[{\hat{A} + \hat{A}^{\dagger}}\bigl]{\hat{U}^{\eta}}\vert{0}\rangle\}$. Here, $\epsilon$, $\eta$, $\eta_{1}$, and $\eta_{2}$ are either zero or $\pi$.  Furthermore, it is easy to show the following: (i) $\Gamma_{I}^{\epsilon, \eta, \eta, \eta} = \Gamma_{I}^{\eta, \eta, \epsilon, \eta} = \Gamma_{I}^{\eta, \eta, \eta, \epsilon} = \Gamma_{I}^{\eta, \epsilon, \eta, \eta}$; (ii) $\Gamma_{I}^{\epsilon, \epsilon, \eta, \eta} = \Gamma_{I}^{\epsilon, \eta, \epsilon, \eta} = \Gamma_{I}^{\epsilon, \eta, \eta, \epsilon}$; (iii) $\Gamma_{P}^{\epsilon, \eta} = \Gamma_{P}^{\eta, \epsilon}$; and (iv) ${\rm Re}\{\langle{0}\vert{\hat{U}^{\epsilon^{\dagger}}}\hat{A}{\hat{U}^{\eta}}\vert{0}\rangle\} = {\rm Re}\{\langle{0}\vert{\hat{U}^{\eta^{\dagger}}}\hat{A}^{\dagger}{\hat{U}^{\epsilon}}\vert{0}\rangle\}$, for $\epsilon$, $\eta \in \{0, \pi\}$. In contrast to Eq. (\ref{EQ:NLSE}), which requires only four quantum circuits to calculate the cost function, Eqs. (\ref{EQ:NLSE1} - \ref{EQ:NLSE3}) require fourteen such circuits to be evaluated on a quantum device. However, Eqs. (\ref{EQ:NLSE1} - \ref{EQ:NLSE3}) enable the determination of the cost function across a broad range of parameter values along a single dimension, rather than at just one point in the parameter space.

\begin{figure}[hbt]\begin{center}
\includegraphics[clip, trim=0.0cm 3.50cm 0.0cm 0.0cm, width=1.00\linewidth, height=1.45\linewidth, angle=0]{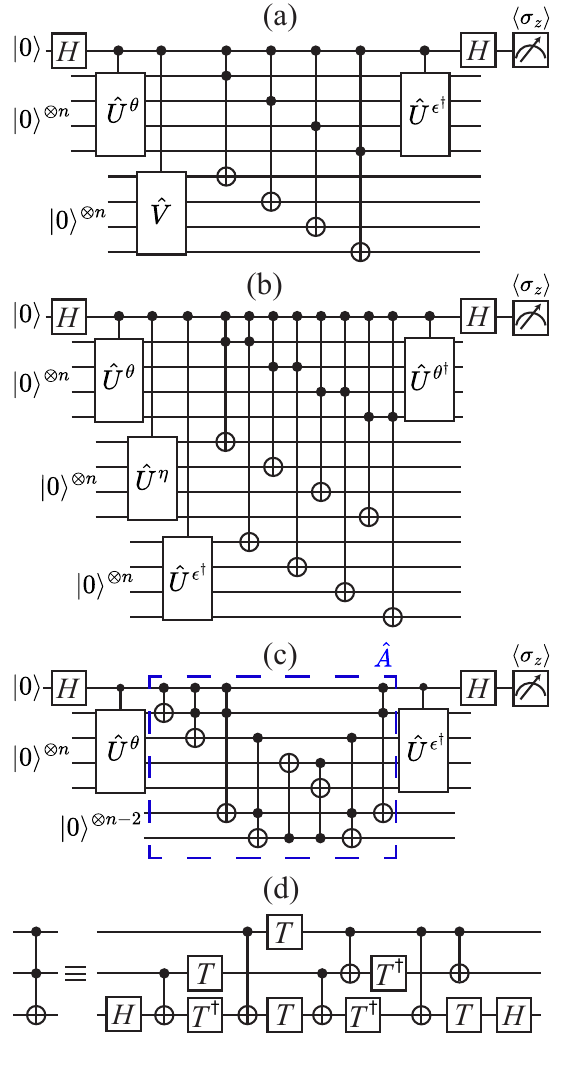}
\caption{Quantum circuits to estimate the components of the (a) potential, (b) interaction, and (c) kinetic energies of the NLSE. Here, $H$ is the Hadamard gate, $\hat{U}^{\lambda_{j_0}}$ is the unitary ansatz operator with parameter $\lambda_{j} = \lambda_{j_{0}}$, while all other variational parameter kept unchanged, $\hat{V}$ is the potential unitary which encodes the potential function values to the basis states. Quantum circuit in panel (a), (b), and (c) estimate the component ${\rm Re} \{\langle{0}\vert\hat{U}^{\epsilon^{\dagger}}\hat{V}\hat{U}^{\theta}\vert{0}\rangle\}$, ${\rm Re} \{\langle{0}\vert\hat{U}^{\theta^{\dagger}}\hat{D}_{2}^{\epsilon^{\dagger}}\hat{D}_{1}^{\eta}\hat{U}^{\theta}\vert{0}\rangle\}$, and ${\rm Re} \{\langle{0}\vert\hat{U}^{\epsilon^{\dagger}}\hat{A}\hat{U}^{\theta}\vert{0}\rangle\}$, respectively.  Here, we have shown an example of $n = 4$, which can be generalized to an arbitrary number of qubits. \vspace{-0.2cm}}
\label{Fig:QNPU_NLSE}
\end{center}\end{figure}

\begin{figure*}[thb]\begin{center}
\includegraphics[clip, trim=0.0cm 0.0cm 0.0cm 0.0cm, width=0.75\linewidth, height=0.9\linewidth, angle=0]{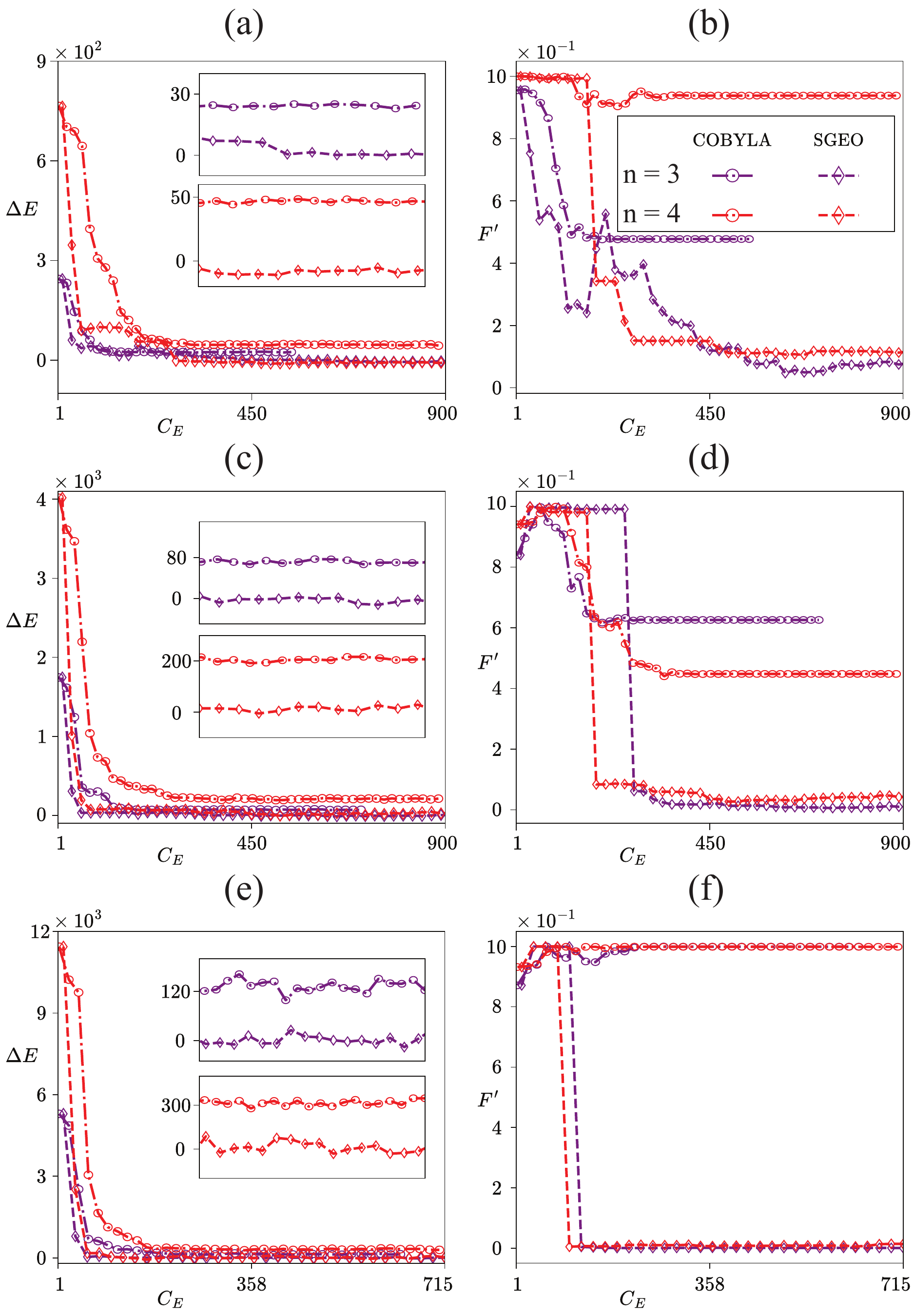}
\caption{Analysis of the convergence of energy difference ($\Delta{E} = \langle\langle{E}\rangle\rangle_{\rm var} - \langle\langle{E}\rangle\rangle_{\rm GS}$) and infidelity ($F' = 1 - \vert\langle\Psi_{\rm target}\vert\Psi_{\rm var}\rangle\vert^{2}$) against the number of circuit evaluations ($C_{E}$) in COBYLA and SGEO processes. We have considered $g = 25$ in (a, b), $g = 250$ in (c, d), and $g = 750$ in (e, f), which represent weak, moderate, and strong regimes of nonlinearity, respectively. Here, purple and red colors indicate the qubit numbers $n = 3$ and $n = 4$, respectively. Moreover, circular and diamond markers represent results obtained using COBYLA and SGEO methods, respectively. \vspace{-0.2cm}}
\label{Fig:NLSE}
\end{center}\end{figure*}

\par Each component of the energy cost function, including the potential, interaction, and kinetic energy, is evaluated individually and requires dedicated quantum circuits, as illustrated in Figs. \ref{Fig:QNPU_NLSE}a-\ref{Fig:QNPU_NLSE}c. For the estimation of the potential (interaction) energy terms ${\rm Re}\{\langle{0}\vert\hat{U}^{\epsilon^{\dagger}}\hat{V}\hat{U}^{\theta}\}$ (${\rm Re}\{\langle{0}\vert\hat{U}^{\theta^{\dagger}}\hat{D}_{2}^{\epsilon^{\dagger}}\hat{D}_{1}^{\eta}\hat{U}^{\theta}\}$) in Eq. (\ref{EQ:NLSE1}) (Eq. (\ref{EQ:NLSE2})), the relevant quantum circuit constructs the potential function (elements of the linear combination of variational states) on separate quantum registers and performs bit-wise multiplication with the state on the primary quantum register, as shown in Fig. \ref{Fig:QNPU_NLSE}a (Fig. \ref{Fig:QNPU_NLSE}b). The elements of the kinetic energy, ${\rm Re}\{\langle{0}\vert\hat{U}^{\epsilon^{\dagger}}\hat{A}\hat{U}^{\eta}\vert{0}\rangle\}$, are computed using an adder circuit \cite{Vedral1996, Nielsen2010, Lubasch2020} that requires an additional $n-2$ ancilla qubits, as depicted in Fig. \ref{Fig:QNPU_NLSE}c, where $n$ denotes the number of qubits in the primary quantum register. Other terms, such as ${\rm Re}\{\langle{0}\vert\hat{U}^{\epsilon^{\dagger}}\hat{A}^{\dagger}\hat{U}^{\eta}\vert{0}\rangle\}$, require identical quantum circuits, where $\hat{A}$ in Fig. \ref{Fig:QNPU_NLSE}c is replaced by $\hat{A}^{\dagger}$. 

\par We now numerically solve the VQA within the domain $x \in [0, 1]$, setting $V_{0} = 1000$, to determine the minimum energy state. In this analysis, we focus on three parameter regimes corresponding to weak, intermediate, and strong nonlinearity, characterized by $g = 25$, $g = 250$, and $g = 750$, respectively. We quantify the behavior of the cost function by evaluating $\Delta{E} = \langle\langle{E}\rangle\rangle_{\rm var} - E_{\rm GS}$, and examine the behavior of the infidelity $F' = 1 - \vert\langle\Psi_{\rm target}\vert\Psi_{\rm var}\rangle\vert^{2}$ of variational state with respect to the exact ground state $\vert\Psi_{\rm target}\rangle$. Here, $\vert\Psi_{\rm target}\rangle$ and $\langle\langle{E_{\rm GS}}\rangle\rangle$ are the exact ground state and corresponding energy value, obtained using the imaginary-time evolution method. The corresponding results are presented in Fig. \ref{Fig:NLSE}, where circular markers represent outcomes of the COBYLA method with $N_{\rm COBYLA} = 300$ iterations, and diamond markers denote our SGEO protocol employing $N_{\rm SGEO} = 10$ iterations. The number of iterations translates to $4N_{\rm COBYLA}$ and $14mN_{\rm SGEO}$ circuit evaluations $C_{E}$ for the COBYLA optimizer and SGEO protocol, respectively, where $m$ denotes the total number of variational parameters. The number of layers in quantum ansatz varies based on the regime of nonlinearity. For intermediate and strong nonlinearity, $d = 3$ and $d = 2$ layers are utilized, respectively. In the weak nonlinearity regime, $d = 2$ layers are used for the 3-qubit system, while $d = 4$ layers are adopted for the 4-qubit system.

\par Across all three parameter regimes, the cost function progressively converges toward its minimum value with increasing optimization iterations or circuit evaluations. The insets in Fig. \ref{Fig:NLSE}a, Fig. \ref{Fig:NLSE}c, and Fig. \ref{Fig:NLSE}e illustrate that the COBYLA optimizer (represented by circular markers) fails to achieve the minimum value, in contrast to our SGEO protocol (depicted by diamond markers), which successfully converges to the global minimum. The cost function notably stagnates after a few iterations of the COBYLA optimizer, failing to reach its minimum value. This observation can be verified by analyzing the infidelity $F'$ of the variational state (circular markers) in Fig. \ref{Fig:NLSE}b, Fig. \ref{Fig:NLSE}d, and Fig. \ref{Fig:NLSE}f, which reveals that the variational state deviates significantly from the exact ground state. This behavior suggests that the COBYLA optimization process may become trapped in local minimum or hindered by barren plateaus. Conversely, our SGEO protocol (diamond markers) effectively identifies the global optimal point, achieving the minimum value of the cost function. This result is corroborated by examining the infidelity of the variational state, as shown in Fig. \ref{Fig:NLSE}b, Fig. \ref{Fig:NLSE}d, and Fig. \ref{Fig:NLSE}f.

\par The analysis presented in this section underscores the efficacy of representing the ansatz as a weighted summation of distinct unitary operators coupled with iterative optimization of the coefficients. Such a strategy not only facilitates the efficient execution of variational algorithms but also mitigates computational burdens while preserving the accuracy of results, even within complex optimization landscapes. Overall, the findings highlight the potential of this method to serve as a versatile and robust framework for enhancing quantum variational algorithms across diverse applications.

\section{ Summary and Outlook }\label{Sec_Conclusion}

\par In this work, we proposed a sequential optimization method, referred to as sequential grid-based explicit optimization, specifically designed for variational quantum algorithms (VQAs). This optimization protocol is based on representing the parameterized quantum circuit (PQC) as a weighted sum of distinct unitary operators, enabling the cost function to be expressed as a sum of multiple terms. This formulation establishes a systematic framework for analyzing variational problems. Notable advantages of this method include the capability to evaluate the cost function over individual parameter domains, $\lambda_{j} \in [-\pi, \pi)$, efficient computation of arbitrary derivatives without additional quantum resources, and explicit parameter optimization using a grid-based evaluation of the cost function. We illustrated the applicability of this method using two distinct cost functions: the first being the squared residual of the variational state relative to a target state, and the second, the expectation value of an observable.

\par Following the development of the general framework, we assessed the efficacy of our method by applying it to two nonlinear physics problems: the dynamics of fluid configurations governed by the one-dimensional Burgers\textquotesingle{} equation and the ground state approximation of the one-dimensional nonlinear Schr\"{o}dinger equation. The results demonstrate that our proposed approach of problem-specific optimization protocol significantly enhances computational efficiency. It consistently outperforms the conventional COBYLA optimizer and exhibits superior convergence properties, even in complex high-dimensional landscapes. These findings underscore the flexibility and robustness of the proposed approach in addressing various optimization challenges within the context of variational quantum algorithms.

\par Future research could comprehensively investigate the potential of problem-tailored optimization processes to effectively navigate complex landscapes of cost functions, which are often populated by local minima traps and barren plateaus. By leveraging the nonlocal characteristics of the cost function and its arbitrary derivatives, it may be possible to develop advanced optimization protocols that successfully navigate the challenges posed by local minima and barren plateaus. Moreover, examining the effects of quantum hardware noise on the efficacy of our approach represents a promising avenue for future study, as it would facilitate the assessment of the optimization process in practical, real-world contexts. Additionally, it is pertinent to explore the decomposition of parameterized quantum circuits that include parameterized multi-qubit operators, thereby offering a systematic methodology for optimizing multi-qubit gates.

\acknowledgments
This research is supported by the National Research Foundation, Singapore and A*STAR under its CQT Bridging Grant, Quantum Engineering Programme NRF2021-QEP2-02-P02, and the EU HORIZON - Project 101080085 - QCFD.

\appendix

\section{ Decomposition of multi-qubit unitary operators }
\label{AppSec:UOperators}

\par In this section, we demonstrate that, in addition to single-qubit operators, certain parameterized two-qubit operators can also be decomposed in a similar manner and analyzed as described in Sec. \ref{Sec:PQC}. First, we consider the parameterized two-qubit gates of the type $\hat{\tilde{U}}(\lambda_{j}) = e^{-i\lambda_{j}\hat{Q}/2} = \cos(\lambda_{j}/2)\hat{I} - i\sin(\lambda_{j}/2)\hat{Q}$. Here, $\hat{Q}$ can be any multi-qubit operator, such as $\hat{Q} \in \{$ $\hat{X}_{j}\hat{X}_{k}$ , $\hat{X}_{j}\hat{X}_{k}\hat{X}_{l}$, $\hat{Y}_{j}\hat{Y}_{k}$, $\hat{Y}_{j}\hat{Y}_{k}\hat{Y}_{l}$, $\hat{Z}_{j}\hat{Z}_{k}$, $\hat{Z}_{j}\hat{Z}_{k}\hat{Z}_{l}$, $\hat{X}_{j}\hat{Y}_{k}$, $\hat{Z}_{j}\hat{X}_{k}$, $\cdots\}$, as long as $\hat{Q}^{2} = \bf{1}$.

\par We now turn our attention to the parameterized SWAP and iSWAP gates. SWAP and iSWAP gates play a significant role in quantum computing, particularly in the manipulation of quantum states and the implementation of quantum algorithms. Their parameterization allows for greater flexibility and control in quantum operations, enhancing the capabilities of quantum circuits. Parameterized SWAP and iSWAP gates are given as,
\bea\bal\label{EQ:swap}
\hat{U}_{\rm (i)SWAP}(\lambda_{j}) &= e^{-i\frac{\lambda_{j}}{2}\bigl[\hat{X}\hat{X} \pm \hat{Y}\hat{Y}\bigl]} = e^{-i\frac{\lambda_{j}}{2}\hat{X}\hat{X}}e^{\mp{i}\frac{\lambda_{j}}{2}\hat{Y}\hat{Y}}\;,~~~
\eal\eea
\vspace{-2.0mm}
\bea\bal\label{EQ:swap1}
\hat{U}_{\rm (i)SWAP}(\lambda_{j}) &= \cos^{2}(\lambda_{j}/2)\hat{I} \pm \sin^{2}(\lambda_{j}/2)\hat{Z}\hat{Z} \\ &~~~ - \frac{i}{2}\sin(\lambda_{j})\bigl[\hat{X}\hat{X} \pm \hat{Y}\hat{Y}\bigl]\;.~~
\eal\eea
From Eq. (\ref{EQ:swap1}), we recognize that $\hat{I} \equiv \hat{U}_{\rm (i)SWAP}(0)$, $\pm\hat{Z}\hat{Z} \equiv \hat{U}_{\rm (i)SWAP}(\pi)$, and $-i\bigl[\hat{X}\hat{X} \pm \hat{Y}\hat{Y}\bigl] \equiv 2\hat{U}_{\rm (i)SWAP}(\pi/2) - \hat{U}_{\rm (i)SWAP}(0) - \hat{U}_{\rm (i)SWAP}(\pi)$. These relations allow us to express the parameterized (i)SWAP gates as, 
\bea\bal\label{EQ:swap2}
\hat{U}_{\rm (i)SWAP}(\lambda_{j}) &= \bigl[\cos^{2}(\lambda_{j}/2) - \frac{\sin(\lambda_{j})}{2}\bigl]\hat{U}^{0}_{\rm (i)SWAP}~~~~\\ 
&~~~~+ \bigl[\sin^{2}(\lambda_{j}/2) - \frac{\sin(\lambda_{j})}{2}\bigl]\hat{U}^{\pi}_{\rm (i)SWAP} ~~~~~~\\
&~~~~+ \sin(\lambda_{j})\hat{U}^{\pi/2}_{\rm (i)SWAP} \;, ~~
\eal\eea
where $\hat{U}^{\lambda_{j_{0}}}_{\rm (i)SWAP} \equiv \hat{U}_{\rm (i)SWAP}(\lambda_{j} = \lambda_{j_{0}})$. With this decomposition, the squared residual (expectation) cost function can be expressed as a sum of three (nine) terms. It is important to note that other parameterized two- or multi-qubit unitary operators can also be represented in a similar fashion.

\section{ Burgers\textquotesingle{} equation cost function } \label{AppSec:Burger}

\par We define square of the residual cost function for the nonlinear dynamics as 
\bea\bal\label{AppEQ:BurgersCost1}
C_{I} &= ~\vert\vert~\Lambda_{t+\tau}\vert\Psi_{t+\tau}({\boldsymbol\lambda})\rangle \\ 
&~- \bigl[\Lambda_{t} + l_{1}\big(\hat{A} + \hat{A}^{\dagger} - 2\hat{I}\bigl) - l_{2}\hat{D}_{t}\bigl(\hat{A} - \hat{A}^{\dagger} \bigl)\bigl]\vert\Psi_{t}\rangle~\vert\vert^{2}\;,~~~~
\eal\eea
where $l_{1} = {\Lambda_{t}\tau\nu}/{2\delta_{x}^{2}}$, $l_{2} = {\vert\Lambda_{t}\vert^{2}\tau}/{2\delta_{x}}$, $\hat{D}_{t} = {\rm diag}(\psi_{n_{x}, t})$, and $\hat{A}$ is an adder operator that translates the basis states by one unit \cite{Lubasch2020, Jaksch2023}. Here, $\vert\Psi_{t+\tau}({\boldsymbol\lambda})\rangle = \hat{U}_{t+\tau}({\boldsymbol\lambda})\vert{0}\rangle$ and $\vert\Psi_{t}\rangle = \hat{U}_{t}\vert{0}\rangle$ is the state at time $t+\tau$ and $t$, respectively. Simplifying the Eq. (\ref{AppEQ:BurgersCost1}) results in 
\bea\bal\label{AppEQ:BurgersCost2}
C_{I} &= \vert\Lambda_{t+\tau}\vert^{2} - 2\Lambda_{t+\tau}\Bigl[\bigl(\Lambda_{t}^{*} - 2l_{1}^{*}\bigl)~{\rm Re}\{\langle{0}\vert \hat{U}_{t}^{\dagger}\hat{U}_{t + \tau}({\boldsymbol\lambda})\vert{0}\rangle\} \\ 
& ~~+ l_{1}^{*}{\rm Re}\{\langle{0}\vert \hat{U}_{t}^{\dagger}\bigl(\hat{A} + \hat{A}^{\dagger}\bigl)\hat{U}_{t + \tau}({\boldsymbol\lambda})\vert{0}\rangle\} \\
& ~~+ l_{2}{\rm Re}\{\langle{0}\vert \hat{U}_{t}^{\dagger}\bigl(\hat{A} - \hat{A}^{\dagger}\bigl){\hat{D}_{t}^{\dagger}}\hat{U}_{t + \tau}({\boldsymbol\lambda})\vert{0}\rangle\}\Bigl]~+~{\rm Const.}\;.
\eal\eea
where Const. is independent of hyper-parameter $\Lambda_{t+\tau}$ and variational parameters ${\boldsymbol\lambda}$. Hyper-parameter $\Lambda_{t}$ of the instance $t$ is already known during the previous optimization iteration. In-fact, we can optimize the cost function with respect to hyper-parameter $\Lambda_{t+\tau}$ by taking $\partial{C_{I}}/\partial{\Lambda_{t+\tau}} = 0$, given $\partial^{2}{C_{I}}/\partial{\Lambda_{t+\tau}^{2}} = 2$ which simplifies the cost function to Eq. (\ref{EQ:Burgers3}) of main text, i.e.,
\bea\bal\label{AppEQ:BurgersCost3}
C_{I}({\boldsymbol\lambda}) &= - \Bigl[\bigl(\Lambda^{*}_{t} - 2l_{1}^{*}\bigl)~{\rm Re}\{\langle{0}\vert \hat{U}_{t}^{\dagger}\hat{U}_{t + \tau}({\boldsymbol\lambda})\vert{0}\rangle\} \\ 
& ~~~~~~~+ l_{1}^{*}{\rm Re}\{\langle{0}\vert \hat{U}_{t}^{\dagger}\bigl(\hat{A} + \hat{A}^{\dagger}\bigl)\hat{U}_{t + \tau}({\boldsymbol\lambda})\vert{0}\rangle\} \\
& ~~~~~~~+ l_{2}{\rm Re}\{\langle{0}\vert \hat{U}_{t}^{\dagger}\bigl(\hat{A} - \hat{A}^{\dagger}\bigl){D_{t}^{\dagger}}\hat{U}_{t + \tau}({\boldsymbol\lambda})\vert{0}\rangle\} \Bigl]^{2}\;,~~~~
\eal\eea
where we have ignored the constant term which only shift the magnitude of the cost function toward zero value.

\bibliography{References_QCFD}

\raggedend
\end{document}